# Foundations of a theory of gravity with a constraint and its canonical quantization

**Alexander P. Sobolev**[a)]

**AFFILIATIONDS**

Moscow Institute of Physics and Technology, Russian Federation

[a)]**Author to whom correspondence should be addressed:** sobolev.ap0439@gmail.com

## ABSTRACT

The gravitational equations were derived in general relativity (GR) using the assumption of their covariance relative to arbitrary transformations of coordinates. It has been repeatedly expressed an opinion over the past century that such equality of all coordinate systems may not correspond to reality. Nevertheless, no actual verification of the necessity of this assumption has been made to date. The paper proposes a theory of gravity with a constraint, the degenerate variants of which are general relativity (GR) and the unimodular theory of gravity. This constraint is interpreted from a physical point of view as a sufficient condition for the adiabaticity of the process of the evolution of the space-time metric. The original equations of the theory of gravity with the constraint are formulated. On this basis, a unified model of the evolution of the modern, early, and very early Universe is constructed that is consistent with the observational astronomical data but does not require the hypotheses of the existence of dark energy, dark matter or inflatons. It is claimed that: physical time is anisotropic, the gravitational field is the main source of energy of the Universe, the maximum global energy density in the Universe was 64 orders of magnitude smaller the Planckian one, and the entropy density is 18 orders of magnitude higher the value predicted by GR. The value of the relative density of neutrinos at the present time and the maximum temperature of matter in the early Universe are calculated. The wave equation of the gravitational field is formulated, its solution is found, and the nonstationary wave function of the very early Universe is constructed. It is shown that the birth of the Universe was random.

## I. INTRODUCTION

Over a hundred years ago, in the derivation of the gravitational equations from the variational principle, Hilbert formulated "an axiom of the general invariance of the action in relation to arbitrary transformations of the world parameters [coordinates]" and chose "$R$ – the invariant built from the Riemann tensor [curvature of the four-dimensional manifold]" as the Lagrangian of the gravitational field.[1]

Three years earlier, Einstein wrote: "Besides, it should be emphasized that we have no basis whatever for assuming general covariance of the gravitational equations …. From this it seems to follow that the equations sought will be covariant only with respect to a particular group of transformations, which for the time being, however, is unknown to us. It seems most natural to demand that the system of equations should be covariant against *arbitrary* [Einstein's italics] transformations." (Ref. 2, pp. 237, 243).

The success of the canonical theory of gravity ostensibly corroborated the validity of such an assumption, and it eventually acquired the status of a fundamental principle, although the opposite point of view had also been expressed earlier (Ref. 3, p. 631): "…the physical meaning of GR [general relativity] boils down to the creation of a new theory of gravity. However, Einstein, the author of the theory, had another point of view, as do a number of his followers. They believe that in addition to this, and fundamentally, GR establishes the principle of the equality of all reference frames. It is difficult to agree with this position, however, since this illegitimately interprets the equality of reference frames from the perspective of a formal mathematical apparatus as equality in terms of their physical essence."

In the light of new experimental data, GR no longer seems as unshakeable as it once did.[4–6] For an explanation of the results derived within the framework of this theory, it was necessary to introduce certain hypothetical entities (the ΛCDM model[7]) the nature of which are still unclear. "*Entia non sunt multiplicanda praeter necessitatem"*; it is likely that the necessity for the introduction of inflatons at first, and now of dark energy and dark matter in GR (with the development of new methods of astronomical observation), are symptoms of a defect in its fundamental basis.

General relativity violates the material unity of the world. In GR, the gravitational field itself does not have the properties of a material medium; its energy–momentum density is zero. This is a direct consequence of the general covariance of the gravitational field equations. Attempts to introduce a non-general covariant energy–momentum density actually mean refuting the original axiom of general covariance.

In my opinion, *it is the general covariance of the equations that is the source of the troubles of GR*.

One possible way to construct a *non-generally covariant theory of gravity without violating Hilbert's axioms* (as I see it) is the introduction of an *a priori* constraint that restricts the choice of coordinate system. Attempts of such a kind have been made previously, for example the unimodular theory of gravity, whose origins date back to Einstein. A consequence of the introduction of this constraint is the appearance of an edge in the space–time manifold. Therefore, *restrictedly covariant geometric objects are defined only on manifolds with this edge*.

Under such an approach, the fundamental principle of the equivalence of all reference systems compatible with the pseudo-Riemannian metric, which underlies GR, is not violated. In addition, we do not put into doubt the principle of the invariance of *matter* action relative to arbitrary transformations of coordinates. At the same time, in contrast to GR, *a covariance of the gravitational equations is restricted by the constraint*. Thus, *a priori*, only the "medium-strong principle" of equivalence is met in this case.[8] However, this cannot be grounds for rejecting the proposed approach as contradicting the experiments verifying the strong equivalence principle for bodies of cosmic scales.[9]

The fact is that already in GR, within the framework of the ΛCDM model, space itself is endowed with energy. The same thing occurs when an *a priori* constraint is introduced. Space becomes a self-gravitating object because of the nonlinearity of the gravitational equations. One can determine the inertial and gravitational masses of such an object. The solution of the gravitational equations has enough free parameters to not only ensure the requirement of the equality of the inertial mass of the gravitational field to its gravitational mass, but also to determine inertial mass in accordance with Mach's principle (the latter problem has not been solved in GR). From this point of view, the results of experiments[9] should be considered as an indication that *only such (quasi) stationary self-gravitating objects exist for which inertial mass is equal to gravitational mass*.

Hilbert's axioms are formulated in a coordinate language. The gravitational field was represented by the ten components $g_{\mu\nu}(x^\lambda)$ of the metric tensor. In addition, it was assumed that derivatives of the metrics no higher than second order could enter into the gravitational equations.

There is no theorem prohibiting the existence of a constraint between the components of a metric in mathematical physics. However, the unimodular theory turned out to be unacceptable from a physical point of view, which prompted Einstein to abandon it in favor of the general covariant theory. Currently, such theories are considered as an approach to the construction of a quantum theory of gravity.[10] Among the other possible approaches, *a restriction of general covariance has the least effect on the concepts about the world around us that are dictated by common sense*.[11] Of course, there must be sufficiently substantial physical grounds to introduce the restrictions on the group of coordinate transformations.

There is a deep analogy between the mathematical description of gravitational interaction in GR and the description of gauge interactions in elementary particle physics.[12] The only way to calibrate for the latter (due to the requirement for general covariance) is by imposing the condition that the 4-divergence of the gauge fields is equal to zero. A similar condition for the gravitational field would be the requirement for an equality to zero of a 4-divergence of the connectivity consistent with the metric, simplified by a pair of indices $\Gamma^{\rho}_{\nu\rho}$. However, due to the fact that GR is not a gauge theory,[13] to avoid contradictions with the initial provisions, such a condition should be considered not as a gauge, but as a constraint. This constraint must be resolved before obtaining the motion equations from the variational principle.

My basic assumption is that (in contrast to the unimodular theories) *the components of the metric tensor* $g_{\mu\nu}(x^{\lambda})$ *are constrained by the following conservation law* (the physical interpretation of this constraint is given in Sec. II):

$$\frac{\partial}{\partial x^{\mu}}\left(\sqrt{-g}\,g^{\mu\nu}\Gamma^{\rho}_{\nu\rho}\right) \equiv \frac{\partial}{\partial x^{\mu}}\left(g^{\mu\nu}\frac{\partial\sqrt{-g}}{\partial x^{\nu}}\right) = 0,\ \ g = \det\left(g_{\mu\nu}\right),\ g^{\mu\nu}g_{\nu\lambda} = \delta^{\mu}_{\lambda}\ (\mu,\nu = 0, 1, 2, 3). \quad (1.1)$$

The left-hand side of (1.1) is not a generally covariant scalar. For an arbitrary coordinate transformation $x^{\mu} \to x'^{\mu}$,[8]

$$g'(x') = g(x) \times J^2, \quad J = \det\left(\frac{\partial x^{\mu}}{\partial x'^{\nu}}\right). \quad (1.2)$$

It follows from the definition of a scalar that the determinant of the metric tensor changes as a scalar under transformations of coordinates with the Jacobian of the transformation equal to unity in modulus

$$g'(x') = g(x), \quad J = \pm 1. \quad (1.3)$$

On the restricted group of coordinate transformations on which $\sqrt{-g}$ is a scalar, the constraint becomes a geometrical object in some region of the space–time continuum and acquires a physical meaning.

Thus, *the constraint* (1.1) *is a geometric object and defines an edge of the manifold only at the restriction of the group of admissible coordinate transformations, from local general diffeomorphisms to special diffeomorphisms with the Jacobian equal to unity.* In addition, *the constraint* (1.1) *allows global linear transformations of coordinates.*

## II. GRAVITATIONAL FIELD EQUATIONS IN THE PRESENCE OF THE CONSTRAINT

In the currently accepted notations, the Hilbert action has the form

$$S_{gr} = -\frac{c^3}{16\pi G}\int R\sqrt{-g}\,d^4x,$$

where $R = g^{\mu\nu}R_{\mu\nu}$ is the scalar curvature, $R_{\mu\nu}$ is the Ricci tensor,

$$R_{\mu\nu} = \frac{\partial}{\partial x^{\lambda}}\Gamma^{\lambda}_{\mu\nu} - \frac{\partial}{\partial x^{\nu}}\Gamma^{\lambda}_{\mu\lambda} + \Gamma^{\lambda}_{\mu\nu}\Gamma^{\rho}_{\lambda\rho} - \Gamma^{\lambda}_{\mu\rho}\Gamma^{\rho}_{\nu\lambda},$$

and $\Gamma^{\lambda}_{\mu\nu}$ is the Christoffel symbols,

$$\Gamma^{\lambda}_{\mu\nu} = \frac{1}{2}g^{\lambda\rho}\left(-\frac{\partial g_{\mu\nu}}{\partial x^{\rho}} + \frac{\partial g_{\rho\mu}}{\partial x^{\nu}} + \frac{\partial g_{\nu\rho}}{\partial x^{\mu}}\right).$$

The derivation of the gravitational field equations from the Hilbert action in the presence of the constraint is a variational problem involving a conditional extremum. The standard method for solving such problems in cases where the constraints are not solvable in an explicit form is the method of Lagrange multipliers. Introducing a Lagrange multiplier (the scalar field $\Phi$), we write the action in the presence of the constraint (1.1) in the form

$$S_{gr} = -\frac{c^3}{16\pi G}\int (R+Q)\sqrt{-g}d^4x,\; Q = \frac{1}{\sqrt{-g}}\frac{\partial\sqrt{-g}}{\partial x^\mu}g^{\mu\nu}\frac{\partial\Phi}{\partial x^\nu}. \tag{2.1}$$

Since $Q$ is a restrictedly covariant scalar, integration is defined not on a manifold but only on a manifold with an edge, unlike the Hilbert action. Now *all the components of the metric tensor and the scalar $\Phi$ can be considered as independent quantities*, and when the action is varied, we obtain an equation that determines the edge, along with the equations of motion.

When varying the action with respect to field $\Phi$ (instead of the equals sign, the arrow indicates that the full derivatives that do not contribute to the equations of motion are omitted), we obtain

$$\delta S_{gr} = -\frac{c^3}{16\pi G}\int \delta(Q\sqrt{-g})d^4x = -\frac{c^3}{16\pi G}\int \left(\frac{\partial\sqrt{-g}}{\partial x^\mu}g^{\mu\nu}\frac{\partial\delta\Phi}{\partial x^\nu}\right)d^4x \to$$
$$\frac{c^3}{16\pi G}\int \delta\Phi\frac{\partial}{\partial x^\nu}\left(g^{\mu\nu}\frac{\partial\sqrt{-g}}{\partial x^\mu}\right)d^4x.$$

From the principle of stationary action, in view of the arbitrariness of the $\Phi$ variation, we derive Eq. (1.1).

The scalar curvature is covariant relative to arbitrary coordinate transformations; therefore, the calculation of its variation, and accordingly its contribution to the field equations, does not differ from that in Ref. 8.

The presence in the Lagrangian of the additional members besides the scalar curvature gives a contribution at the metric variation

$$\delta(Q\sqrt{-g}) = \left[\frac{\partial\delta\sqrt{-g}}{\partial x^\mu}g^{\mu\nu}\frac{\partial\Phi}{\partial x^\nu} + \frac{\partial\sqrt{-g}}{\partial x^\mu}(\delta g^{\mu\nu})\frac{\partial\Phi}{\partial x^\nu}\right] \to$$
$$\frac{1}{2}\left[g_{\mu\nu}\frac{\partial}{\partial x^\rho}\left(g^{\rho\lambda}\frac{\partial\Phi}{\partial x^\lambda}\right) + \frac{1}{\sqrt{-g}}\frac{\partial\sqrt{-g}}{\partial x^\mu}\frac{\partial\Phi}{\partial x^\nu} + \frac{1}{\sqrt{-g}}\frac{\partial\sqrt{-g}}{\partial x^\nu}\frac{\partial\Phi}{\partial x^\mu}\right]\sqrt{-g}\delta g^{\mu\nu}.$$

This leads to the occurrence of a new object $(\varepsilon_{gr})_{\mu\nu}$ in the Hilbert–Einstein equations along with the energy–momentum tensor of matter $(\varepsilon_{mat})_{\mu\nu}$:

$$R_{\mu\nu} - \frac{1}{2}g_{\mu\nu}R = \frac{8\pi G}{c^4}(\varepsilon_{gr})_{\mu\nu} + \frac{8\pi G}{c^4}(\varepsilon_{mat})_{\mu\nu}, \tag{2.2}$$

$$\frac{8\pi G}{c^4}(\varepsilon_{gr})_{\mu\nu} = -\frac{1}{2}\left[g_{\mu\nu}\frac{\partial}{\partial x^\rho}\left(g^{\rho\lambda}\frac{\partial\Phi}{\partial x^\lambda}\right) + \frac{1}{\sqrt{-g}}\frac{\partial\sqrt{-g}}{\partial x^\mu}\frac{\partial\Phi}{\partial x^\nu} + \frac{1}{\sqrt{-g}}\frac{\partial\sqrt{-g}}{\partial x^\nu}\frac{\partial\Phi}{\partial x^\mu}\right]. \tag{2.3}$$

Object (2.3) contains ordinary derivatives instead of covariant ones and therefore behaves like a tensor only under a restricted group of coordinate transformations. It is *covariant only relative to local special diffeomorphisms and global linear transformations of coordinates*. Since the remaining terms in (2.2) are generally covariant, on the whole, the system of gravitation equations will be covariant only relative to the indicated restricted group of coordinate transformations in the presence of the constraint.

Since the covariant derivative is defined for arbitrary coordinate transformations, its action is also defined for objects that are tensors relative to restricted group of transformations. The only difference is that the new objects belong again to the same type of tensors on which it acts.

Constraint (1.1) does not include matter fields. Therefore, the action for matter remains invariant under general coordinate transformations, as in GR. The covariant derivative of the expression on the left-hand side of (2.2) is zero for mixed tensors in view of the reduced Bianchi identity (the validity of which is due only to general covariance of the curvature tensor); therefore, taking into account the above, the derivative of the sum on the right-hand side of (2.2) must also be equal to zero. Thus, the object $(\varepsilon_{gr})_{\mu\nu}$ changes as a tensor at the stated transformations of coordinates, is symmetric, is a source of curvature of space–time like matter, and in the absence of matter, its covariant derivative on the field equation is equal to zero.

All this in aggregate makes it possible to call object (2.3) an energy–momentum density tensor of the gravitational field, expressed using the auxiliary field $\Phi(x^\mu)$. The field is auxiliary because it does not initially enter either the Hilbert action, the matter action, or the constraint equation. At the same time, the introduction of the field $\Phi(x^\mu)$ is inevitable in the very essence of the mathematical problem. It is also impossible, without conflicting with the principles of the calculus of variations, to give the field $\Phi(x^\mu)$ a physical meaning, simply by subordinating it to some generally covariant field equation. Section III consider the case where it is possible to explicitly exclude the field $\Phi(x^\mu)$ from the gravitational field equations. The question of the positive definiteness of the energy density of the gravitational field will also be considered there.

Thus, we have derived the system of equations involving constraint (1.1) and ten equations (2.2) for eleven unknowns listed above.

From a physical point of view, constraint (1.1) can be interpreted as a sufficient condition for adiabaticity of the metric evolution process. We determine the vector of *the entropy density flux of the gravitational field* by the relation

$$s_{gr}v^\mu = \text{const} \times g^{\mu\lambda}\frac{\partial \ln\sqrt{-g}}{\partial x^\lambda}, \quad v^\mu v_\mu = 1, \quad s_{gr} = \text{const} \times v^\lambda(x)\frac{\partial \ln\sqrt{-g}}{\partial x^\lambda}. \qquad (2.4)$$

In the Planck system of units, this constant can be written in the form

$$\text{const} = a \times \frac{k}{l_{pl}^2}, \quad l_{pl}^2 = \frac{\hbar G}{c^3}, \qquad (2.5)$$

where $k$ is the Boltzmann constant and $l_{pl}$ is the Planck length. For a quasi-classical theory, the condition $|a| \leq 1$ must be satisfied. The sign of the constant $a$ must be chosen so that the entropy density would be positive on time-like geodesic lines. Now constraint (1.1) can be written in the form of the relativistic adiabaticity condition [14]

$$\frac{\partial}{\partial x^\mu}\left(\sqrt{-g}s_{gr}v^\mu\right) = 0. \qquad (2.6)$$

We note that under definition (2.4), all the thermodynamic potentials will be scalars only relative to the restricted group of transformations.

Thus, *the determination of the energy densities* (2.3) *and entropy* (2.4) *satisfying the conservation laws allows us to consider the gravitational field as an ordinary material medium*[14] *and restores the material unity of the world violated by GR.* In order for these definitions to become meaningful, it is necessary to impose restrictions on the quantities included in them.

In the case when $\Phi(x)$ = const, it follows from the definition (2.3) that $(\varepsilon_{gr})_\mu^\nu = 0$ and the system of equations (2.2) goes over into the generally covariant system of equations of GR with all its inherent conceptual problems. In this case, the constraint (1.1) partially reduces the ambiguity in a definition of the metric tensor. For its complete definition, one can impose the condition $g_{0m}(x^\mu) = 0$ ($m$ = 1, 2, 3), which allows synchronization of clocks at different points in space.[15]

The energy - momentum density of the gravitational field will differ from zero only under the condition $\Phi(x) \neq$ const. In this case, the system of equations (2.2) will be restrictedly covariant in contrast to GR. In particular, this includes theories with a metric whose determinant is constant. However, this option is also unacceptable from a physical point of view, since the entropy density of the gravitational field (2.4) is equal to zero in this case.

Thus, the gravitational field will actually have all the properties of the material medium only if two conditions are met:

$$\Phi(x) \neq \text{const}, \;\; g(x) \neq \text{const}. \qquad (2.7)$$

In GR, the first of these conditions is violated, and in the unimodular theory of gravity, the second is violated. In what follows, we will consider the conditions (2.7) fulfilled.

For inclusion in the consideration of spinor matter and gauge fields, the system of equations (1.1, 2.2) can be formulated in a nonholonomic orthogonal frame. In addition to this, along with the affine connection, the spin connection is introduced.[8] This is possible, despite the presence of the constraint, since the group of local Lorentz transformations is unimodular.

## III. EVOLUTION OF THE SPACE–TIME MANIFOLD IN THE ABSENCE OF MATTER

If we imagine that matter and the radiations generated by it were absent at the initial instant of time in the Universe, there would not be physical possibility to distinguish the points of outer space. What could be the metric properties of such a space in this case?

There are nine possible types of principal homogeneous spaces (admitting a group of motions) with a time-dependent metric (the Bianchi classification) in three-dimensional space.[15] The introduction of the constraint restricts not only the group of coordinate transformations admissible in GR, but also the group of motions that preserve the metric. If the first group is given by condition (1.3), then at motion, by virtue of the requirement of form-invariance, this condition takes the form

$$g(x^0, x'^m) = g(x^0, x^m), \quad x^m \to x'^m \quad (m = 1,\ 2,\ 3),$$

that is, the determinant of the metric tensor does not depend on spatial coordinates. Note that this does not exclude the dependence of the components of the metric tensor on coordinates. These dependences are given in Ref. 16 (Ch. V, § 31) for all nine types of homogeneous spaces. Calculating the determinant of the metric tensor, we make sure that it does not depend on spatial coordinates only for homogeneous spaces of type I and II according to the Bianchi classification. This means that if constraint (1.1) applies, only these two types of homogeneous spaces can exist:

$$\text{I. } g_{mn} = a_{mn}(x^0),\ g_{00} = a_{00}(x^0) > 0,\ g_{0n} = 0 \quad (m, n = 1,\ 2,\ 3),$$

$$\text{II. } g_{mn} = \begin{pmatrix} a_{11} & a_{12} & a_{12}x^1 + a_{13} \\ a_{12} & a_{22} & a_{22}x^1 + a_{23} \\ a_{12}x^1 + a_{13} & a_{22}x^1 + a_{23} & a_{22}(x^1)^2 + 2a_{23}x^1 + a_{33} \end{pmatrix},$$

$$a_{mn} = a_{mn}(x^0),\ g_{00} = a_{00}(x^0),\ g_{0n} = 0.$$

For the first of these, the components of the metric tensor depend only on time. In this case, if the spatial metric is non-degenerate, then the most general expression for the space–time interval is the transformation of coordinates with the Jacobian equal to unity,[15]

$$x^0 \to x^0, \quad x^m \to x^m + \phi^m(x^0),$$

which can always be reduced to the form

$$ds^2 = g_{00}(x^0)(dx^0)^2 + g_{mn}(x^0)dx^m dx^n,\ \gamma = -\det(g_{mn}) > 0 \ \ (m, n = 1,\ 2,\ 3). \quad (3.1)$$

### A. Gravitational equations for homogeneous spaces of type I

The absence of general invariance of action (2.1) does not allow us to eliminate the metric component $g_{00}$. The expressions for the Christoffel symbols and the nonzero components of the Ricci tensor for metric (3.1) will take the form

$$\Gamma^0_{00} = \frac{1}{2}g^{00}\frac{dg_{00}}{dx^0},\ \Gamma^0_{0l} = 0,\ \Gamma^0_{nl} = -\frac{1}{2}g^{00}\frac{dg_{nl}}{dx^0},\ \Gamma^m_{00} = 0,\ \Gamma^m_{0l} = \frac{1}{2}g^{mk}\frac{dg_{kl}}{dx^0},\ \Gamma^m_{nl} = 0, \quad (3.2)$$

$$R^0_0 = -\frac{1}{2\sqrt{g_{00}}}\frac{d}{dx^0}\left(\frac{1}{\gamma\sqrt{g_{00}}}\frac{d\gamma}{dx^0}\right) - \frac{1}{4g_{00}}g^{mk}\frac{dg_{kp}}{dx^0}g^{pn}\frac{dg_{nm}}{dx^0}, \quad (3.3)$$

$$R^p_k = -\frac{1}{2\sqrt{\gamma g_{00}}}\frac{d}{dx^0}\left(\sqrt{\frac{\gamma}{g_{00}}}g^{mp}\frac{dg_{km}}{dx^0}\right), \quad (3.4)$$

and the nonzero components of the energy–momentum density tensor (2.3) for metric (3.1) will take the form

$$\left(\varepsilon_{gr}\right)^0_0 = -\frac{c^4}{16\pi G}\left[\frac{d}{dx^0}\left(\frac{1}{g_{00}}\frac{d\Phi}{dx^0}\right) + \frac{2}{g_{00}\sqrt{g_{00}\gamma}}\frac{d\sqrt{g_{00}\gamma}}{dx^0}\frac{d\Phi}{dx^0}\right], \quad (3.5)$$

$$\left(\varepsilon_{gr}\right)^p_k = -\frac{c^4}{16\pi G}\frac{d}{dx^0}\left(\frac{1}{g_{00}}\frac{d\Phi}{dx^0}\right)\delta^p_k. \quad (3.6)$$

Taking these relations into account, the gravitational field equations (2.2) in mixed components

$$R^\lambda_\mu = \frac{8\pi G}{c^4}\left[(\varepsilon_{gr})^\lambda_\mu - \frac{1}{2}\delta^\lambda_\mu(\varepsilon_{gr})^\nu_\nu\right]$$

in the presence of the constraint will take the form

$$\frac{d}{dx^0}\left(\frac{1}{g_{00}}\frac{d\sqrt{\gamma g_{00}}}{dx^0}\right) = 0, \tag{3.7}$$

$$-\frac{1}{2\sqrt{g_{00}}}\frac{d}{dx^0}\left(\frac{1}{\gamma\sqrt{g_{00}}}\frac{d\gamma}{dx^0}\right) - \frac{1}{4g_{00}}g^{mk}\frac{dg_{kp}}{dx^0}g^{pn}\frac{dg_{nm}}{dx^0} = \frac{\sqrt{\gamma g_{00}}}{2}\frac{d}{dx^0}\left(\frac{1}{g_{00}\sqrt{\gamma g_{00}}}\frac{d\Phi}{dx^0}\right), \tag{3.8}$$

$$-\frac{d}{dx^0}\left(\sqrt{\frac{\gamma}{g_{00}}}g^{mp}\frac{dg_{km}}{dx^0}\right) = \delta_k^p\frac{d}{dx^0}\left(\sqrt{\frac{\gamma}{g_{00}}}\frac{d\Phi}{dx^0}\right). \tag{3.9}$$

**B. Solution of the system of equations (3.7 … 3.9)**

The system of equations (3.7 … 3.9) is a nonlinear system of eight equations for eight unknown functions of the world coordinate time: $\Phi(x^0)$, $g_{00}(x^0)$, $g_{mn}(x^0)$ ($m$, $n$ = 1, 2, 3). We show that there is an exact general solution to this nonlinear system of equations.

Eq. (3.9) shows that

$$g^{mp}\frac{dg_{km}}{dx^0} + \delta_k^p\frac{d\Phi}{dx^0} = \sqrt{\frac{g_{00}}{\gamma}}L_k^p. \tag{3.10}$$

The constant matrix $L_k^p$ is not arbitrary. Since Eq. (3.10) shows that

$$\frac{dg_{kn}}{dx^0} + g_{kn}\frac{d\Phi}{dx^0} = \sqrt{\frac{g_{00}}{\gamma}}g_{np}L_k^p, \tag{3.11}$$

the matrix must satisfy the condition

$$g_{np}(x^0)L_k^p \equiv g_{kp}(x^0)L_n^p. \tag{3.12}$$

For a general metric tensor, this condition will be satisfied only in the case where the matrix $L_k^p$ is proportional to the identity matrix. Otherwise, the matrix $L_k^p = \mathrm{diag}(L_1,\ L_2,\ L_3)$, and the metric tensor must also be diagonal.

Simplifying Eq. (3.10) with respect to the indices $p$ and $k$ gives

$$3\frac{d\Phi}{dx^0} = -\frac{1}{\gamma}\frac{d\gamma}{dx^0} + \sqrt{\frac{g_{00}}{\gamma}}L_k^k. \tag{3.13}$$

Thus, in the case of a homogeneous space of type I, it is possible to express *in the explicit form* the derivative of the field $\Phi$ in terms of the metric field and its derivatives. This demonstrates the auxiliary nature of this field. Indeed, it is enough to substitute Eq. (3.13) into the system of equations (3.7 ... 3.9) to obtain seven equations for seven components of a metric.

Substituting (3.13) into (3.10), we get

$$g^{pm}\frac{dg_{km}}{dx^0} = \frac{1}{3\gamma}\frac{d\gamma}{dx^0}\delta_k^p + \sqrt{\frac{g_{00}}{\gamma}}\left(L_k^p - \frac{1}{3}\delta_k^p L_n^n\right). \tag{3.14}$$

Equation (3.14) shows that

$$g^{mk}\frac{dg_{kp}}{dx^0}g^{pn}\frac{dg_{nm}}{dx^0} = \frac{1}{3}\left(\frac{1}{\gamma}\frac{d\gamma}{dx^0}\right)^2 + \frac{g_{00}}{\gamma}\left[L_k^p L_p^k - \frac{1}{3}(L_n^n)^2\right]. \tag{3.15}$$

Using this expression and Eq. (3.13), it is possible to eliminate $\Phi$ and all spatial metric components from Eq. (3.8), and we can write it in the form

$$3\frac{d}{dt}\left(\frac{1}{\gamma}\frac{d\gamma}{dt}\right) + \frac{1}{2}\left(\frac{1}{\gamma}\frac{d\gamma}{dt}\right)^2 + \frac{3c^2}{2\gamma}\left[L_k^p L_p^k - \frac{1}{3}(L_n^n)^2\right] = g_{00}\sqrt{\gamma}\frac{d}{dt}\frac{1}{\gamma g_{00}}\left(\frac{1}{\sqrt{\gamma}}\frac{d\gamma}{dt} - cL_n^n\right), \tag{3.16}$$

where the notation $cdt = \sqrt{g_{00}}dx^0$ is introduced.

Equation (3.7) implies

$$\frac{1}{g_{00}}\frac{dg_{00}}{dt} + \frac{1}{\gamma}\frac{d\gamma}{dt} = \frac{1}{T\sqrt{\gamma}},\ T = \mathrm{const}. \tag{3.17}$$

This equation allows us to eliminate $g_{00}$ from (3.16) and to write the equation for the function $\gamma$:

$$2\frac{d}{d\tau}\left(\frac{1}{\gamma}\frac{d\gamma}{d\tau}\right) + \frac{1}{\gamma\sqrt{\gamma}}\frac{d\gamma}{d\tau} - \frac{\sigma}{\gamma} = 0,\ \sigma = B_n^n - \frac{3}{2}\left[B_k^p B_p^k - \frac{1}{3}(B_n^n)^2\right], \tag{3.18}$$

where $\tau = t/T$ is the dimensionless proper time and $B_k^p = cTL_k^p$ is a matrix of dimensionless constants. The order of Eq. (3.18) can be reduced by introducing the function $u(\gamma)$, which is the dimensionless rate of change of the *volume factor* $\sqrt{\gamma}$

$$u = \frac{d\sqrt{\gamma}}{d\tau}. \tag{3.19}$$

The equation then takes the form

$$8\gamma u\frac{du}{d\gamma} = 4u^2 - 2u + \sigma, \quad \frac{16udu}{(4u-1)^2+4\sigma-1} = \frac{d\sqrt{\gamma}}{\sqrt{\gamma}}. \tag{3.20}$$

Equation (3.20) has no singularities at $\sigma > 1/4$. So that the metric does not have singular points, we will further consider this restriction on the value of σ to be fulfilled.

Integrating Eq. (3.20), we find that

$$\sqrt{\frac{\gamma}{\gamma_{\min}}} = f(u), \quad f(u) = \sqrt{\frac{4u^2-2u+\sigma}{\sigma}}\exp\left[\frac{1}{\sqrt{4\sigma-1}}\left(\text{arctg}\frac{4u-1}{\sqrt{4\sigma-1}} + \text{arctg}\frac{1}{\sqrt{4\sigma-1}}\right)\right], \tag{3.21}$$

where $\sqrt{\gamma_{\min}}$ is *the minimum value of* $\sqrt{\gamma\,(u)}$ *at* $u = 0$.

Differentiating (3.21) with respect to $\tau$ gives

$$\frac{1}{\sqrt{\gamma_{\min}}}\frac{d\sqrt{\gamma}}{d\tau} = \frac{df(u)}{du}\frac{du}{d\tau}, \quad \frac{df}{du} = \frac{4u}{4u^2-2u+\sigma}f(u).$$

Hence, we find the solution of Eq. (3.18) in the parametric form in consideration of (3.19) and (3.21):

$$\tau - \tau_{st} = \sqrt{\gamma_{\min}}\int_0^u \frac{4f(y)}{4y^2-2y+\sigma}dy. \tag{3.22}$$

*Evolution of space begins at the time point* $\tau_{st}$ *from a state of rest with the minimal volume factor.*

From Eq. (3.17), taking into account (3.19), it follows that

$$d\ln(\gamma g_{00}) = \frac{dt}{T\sqrt{\gamma}} = \frac{4du}{4u^2-2u+\sigma}.$$

Integrating this equation,

$$\frac{\gamma(u)g_{00}(u)}{\gamma_{\min}g_{00}(0)} = \exp\left(\int_0^u \frac{4du}{4u^2-2u+\sigma}\right) = \exp\left[\frac{4}{\sqrt{4\sigma-1}}\left(\text{arctg}\frac{4u-1}{\sqrt{4\sigma-1}} + \text{arctg}\frac{1}{\sqrt{4\sigma-1}}\right)\right],$$

and taking into account determination (3.21), we get

$$\sqrt{\frac{g_{00}(u)}{g_{00}(0)}} = \frac{\sigma\cdot f(u)}{4u^2-2u+\sigma}.$$

Using this relation, proceeding from determination (3.19), we can show that

$$\sqrt{g_{00}(x^0)}dx^0 = cT\sqrt{\gamma_{\min}}\frac{4f(u)}{4u^2-2u+\sigma}du. \tag{3.23}$$

The world coordinate time $x^0$ has been determined up to an arbitrary linear transformation. The quantity $u \geq 0$ by definition and does not change under such a transformation. Therefore, the parameter $u$ with the dimensional factor can be called world physical anisotropic time.

**C. Energy–momentum density and scalar curvature of a homogeneous space on the field equations**

Using relations (3.13) and (3.17), we can transform (3.5) as follows:

$$(\varepsilon_{gr})_0^0 = \rho_{gr} = \frac{c^2}{48\pi GT^2}\left[\frac{d}{d\tau}\left(\frac{1}{\gamma}\frac{d\gamma}{d\tau}\right) + \frac{1}{2}\left(\frac{1}{\gamma}\frac{d\gamma}{d\tau}\right)^2 + \frac{1}{2\sqrt{\gamma}\gamma}\frac{d\gamma}{d\tau} - \frac{1}{2\gamma}B_k^k\right]. \tag{3.24}$$

Using Eq. (3.18), we eliminate the second derivative, then

$$\rho_{gr} = \frac{c^2}{96\pi GT^2}\left[\left(\frac{1}{\gamma}\frac{d\gamma}{d\tau}\right)^2 - \frac{3}{2\gamma}[B_k^pB_p^k - \frac{1}{3}(B_k^k)^2]\right] = \frac{c^2}{24\pi GT^2\gamma}\left[u^2 - \frac{3}{8}[B_k^pB_p^k - \frac{1}{3}(B_k^k)^2]\right]. \tag{3.25}$$

The first term in the brackets vanishes at small values of $u$, and the second term characterizing the global anisotropy of space is constant and positive and enters into the expression for the energy density with a minus sign. Now we can answer the question posed in 1972: *"*Accepting the agreement with observations, we want to understand *why the laws of physics should demand (rather than merely* *permit) a universe that is homogeneous and isotropic to high accuracy on large scales* [authors' italics]." (Ref. 17, 30.1, p. 800). *The energy density of the gravitational field will be nonnegative only in the case when a homogeneous space is isotropic* ($B_m^n \propto \delta_m^n$).

In this case, the solution of (3.14) can be presented in the form

$$g_{kn} = \left(\frac{\gamma}{\gamma_{\min}}\right)^{1/3} g_{kn}(0). \tag{3.26}$$

Due to the invariance of the theory with respect to global linear transformations of coordinates, the original metric $g_{kn}(0)$ can always be reduced to a diagonal Euclidean form. Then, taking into account relations (3.26), (3.23), and (3.21), interval (3.1) takes the form of a space-time metric *with a maximally symmetric flat subspace* (Ref. 8, Ch. 13):

$$ds^2 = \left(cT\sqrt{\gamma_{\min}}\frac{4f(u)}{4u^2-2u+\sigma}\right)^2 (du)^2 - f^{2/3}(u)dx^m dx^n \delta_{mn}. \tag{3.27}$$

We note that a homogeneous space of type II has an unremovable anisotropy. Therefore, bearing in mind the connection between the positive definiteness of the energy density and the absence of anisotropy, it can be argued that *from a physical point of view, there is no other noncontradictory theory of a three-dimensional homogeneous space besides type I.*

We introduce the Hubble parameter $H$ and the acceleration parameter $q$ (*instead of the deceleration parameter*[8]) according to *the modern representations*:

$$H \equiv \frac{1}{6T\gamma}\frac{d\gamma}{d\tau}, \quad q \equiv 1 + \frac{1}{6H^2T^2}\frac{d}{d\tau}\left(\frac{1}{\gamma}\frac{d\gamma}{d\tau}\right). \tag{3.28}$$

The substitution of these expressions into (3.18) allows us to derive the equation describing change of the acceleration–deceleration eras:

$$q = \frac{3}{4}\left(\frac{\sqrt{\sigma}}{u(\gamma)} - \frac{1}{\sqrt{\sigma}}\right)^2 + 1 - \frac{3}{4\sigma}. \tag{3.29}$$

This implies that two scenarios are possible. When $\sigma > 3/4$, only acceleration ($q > 0$) is possible. When $3/4 > \sigma > 1/4$, a change of eras is possible: acceleration–deceleration–acceleration. The change of eras occurs when the values

$$u_1 = \frac{\sigma}{1+\sqrt{1-4\sigma/3}} > \frac{\sqrt{3}}{4(\sqrt{3}+\sqrt{2})} \approx 0.1376, \;\; u_2 = \frac{\sigma}{1-\sqrt{1-4\sigma/3}} < \frac{\sqrt{3}}{4(\sqrt{3}-\sqrt{2})} \approx 1.3624. \tag{3.30}$$

*The recently discovered change of eras*[4–6] *indicates that the second scenario takes place.*

The maximum value of the deceleration is reached at $u = \sigma$

$$q_{\max} = 1 - \frac{3}{4\sigma} > -2. \tag{3.31}$$

After the onset of the second era of acceleration, $q$ asymptotically approaches unity according to (3.29). The energy density (3.25) of the isotropic gravitational field is related to the Hubble parameter (3.28) by the relation

$$\rho_{gr} = \frac{c^2}{24\pi G T^2 \gamma(u)} u^2 = \frac{3c^2H^2(u)}{8\pi G}. \tag{3.32}$$

Thus, *the energy density of the gravitational field is proportional to the square of the rate of change of the volume factor and is equal to the critical density at any moment of time.*

The Hubble parameter reaches its maximum value during the era of the first acceleration at $u = \sigma/2 < u_1$,

$$H_{\max} = \frac{\sqrt{\sigma}}{6T\sqrt{\gamma_{\min}}}\exp\left(-\frac{\operatorname{arctg}\sqrt{4\sigma-1}}{\sqrt{4\sigma-1}}\right), \tag{3.33}$$

and then monotonously decreases, tending to the constant value

$$H_\infty = \frac{\sqrt{\sigma}}{6T\sqrt{\gamma_{\min}}}\exp\left(-\frac{1}{\sqrt{4\sigma-1}}\left(\operatorname{arctg}\frac{1}{\sqrt{4\sigma-1}} + \frac{\pi}{2}\right)\right). \tag{3.34}$$

From (3.6), the spatial components of the energy–momentum density tensor are equal in the field equations to

$$\left(\varepsilon_{gr}\right)_k^p = \frac{c^2}{48\pi G T^2}\left[\frac{d}{d\tau}\left(\frac{1}{\gamma}\frac{d\gamma}{d\tau}\right) + \frac{1}{2}\left(\frac{1}{\gamma}\frac{d\gamma}{d\tau}\right)^2 - \frac{1}{2\sqrt{\gamma}\gamma}\frac{d\gamma}{d\tau} + \frac{1}{2\gamma}B_n^n\right]\delta_k^p, \quad (3.35)$$

and differ from the expression for the energy density in the sign of the last two members. These components can possess both positive and negative values during evolution. Eliminating the second derivate again by means of Eq. (3.18) and assuming $(\varepsilon_{gr})_m^n = -p_{gr}\delta_m^n$ (as accepted for macroscopic mediums), the gravitational field pressure can be written as

$$p_{gr} = -\frac{c^2}{48\pi G T^2}\frac{2u^2-2u+\sigma}{\gamma(u)}. \quad (3.36)$$

This implies that when $0.25 < \sigma < 0.5$, there is a change of the pressure sign at the following $u$ values:

$$u_3 = \frac{1-\sqrt{1-2\sigma}}{2} > \frac{\sqrt{2}-1}{2\sqrt{2}} \approx 0.146,\ u_4 = \frac{1+\sqrt{1-2\sigma}}{2} < \frac{\sqrt{2}+1}{2\sqrt{2}} \approx 0.8536. \quad (3.37)$$

*The gravitational field has a positive pressure in the interval $u_3 < u < u_4$; in other cases, it has a negative pressure.*

Let us consider the curvature tensor. Substituting relations (3.14) and (3.15) into (3.3) and (3.4), we find the expressions for the curvature tensor on the field equations:

$$R_0^0 = -\frac{1}{2c^2}\frac{d}{dt}\left(\frac{1}{\gamma}\frac{d\gamma}{dt}\right) - \frac{1}{12c^2}\left(\frac{1}{\gamma}\frac{d\gamma}{dt}\right)^2,\ R_k^k = -\frac{1}{2c^2\sqrt{\gamma}}\frac{d}{dt}\left(\frac{1}{\sqrt{\gamma}}\frac{d\gamma}{dt}\right). \quad (3.38)$$

Excluding the second derivatives, we can write the expressions for the scalar curvature of space–time $R$:

$$R = R_0^0 + R_k^k = -\frac{1}{2c^2T^2\gamma(u)}\left(\frac{8}{3}u^2 - 2u + \sigma\right). \quad (3.39)$$

*The space–time curvature changes during evolution and possesses at first negative values, then positive values, and finally negative values once more.* Taking into account (3.32), (3.36), the last relation can be represented as

$$\rho_{gr}(u) - 3p_{gr}(u) + \frac{c^2}{8\pi G}R(u) = 0.$$

**D. Kinematics of a homogeneous space**

According (3.32) and (3.33), the maximum density of the gravitational field energy is

$$\rho_{gr\max} = \frac{c^2\sigma}{96\pi G T^2\gamma_{\min}}\exp\left(-\frac{2\operatorname{arctg}\sqrt{4\sigma-1}}{\sqrt{4\sigma-1}}\right). \quad (3.40)$$

Hence, for $\sigma \approx 1/4$,

$$T\sqrt{\gamma_{\min}} = \left(\frac{c^2\sigma}{96\pi\cdot G\cdot\rho_{gr\max}}\right)^{1/2}\exp\left(-\frac{\operatorname{arctg}\sqrt{4\sigma-1}}{\sqrt{4\sigma-1}}\right) \approx \frac{c}{8e}\left(\frac{1}{6\pi G\rho_{gr\max}}\right)^{1/2}. \quad (3.41)$$

Relations (3.23) and (3.28) can be written in the dimensional form

$$t - t_{st} = T\sqrt{\gamma_{\min}}\int_0^u \frac{4f(y)}{4y^2-2y+\sigma}dy,\ H(u) = \frac{1}{3T\sqrt{\gamma_{\min}}}\frac{u}{f(u)}. \quad (3.42)$$

According (3.21), $f(u)$ depends only on the constant $\sigma$. Substituting the current values (Ref. 18, pp. 110, 111) of the time from the beginning of evolution till now ($t^0 - t_{st} = 13.81 \times 10^9$ years) and the Hubble parameter ($H^0 = 67.3$ km·s$^{-1}$·Mpc$^{-1}$) into these relations gives a pair of equations with two unknowns ($\sigma$ and the value of the parameter $u^0$ at the current time):

$$t^0 - t_{st} = T\sqrt{\gamma_{\min}}\int_0^{u^0} \frac{4f(y)}{4y^2-2y+\sigma}dy,\ H^0 = \frac{1}{3T\sqrt{\gamma_{\min}}}\frac{u^0}{f(u^0)}.$$

The quasi-classical approach is justified providing that the parameter $T\sqrt{\gamma_{\min}} \geq t_{pl}$, where $t_{pl}$ is the Planck time. *According to* (3.40)*, the maximum energy density of the gravitational field, which is four orders of magnitude smaller than the Planck density, corresponds to the minimum value of this parameter*. In this case, the solution of the system of equations is

$\sigma = 0.2501278984$, $u^0 = 6.118625359$.

The results of the calculations of other parameters for this case are presented in Table I. In Table II, the results of a similar calculation are given, but with the maximum energy density equal to that achieved on accelerators with an energy of 1 TeV ($\rho_{gr\max}$ = (1 TeV)$^4$ ≈ 2 × $10^{49}$ J·m$^{-3}$).

The characteristic values $u^0$, $u2$, $u4$, $\sigma$, $u3$, $u1$, and $\sigma/2$, supplemented by a number of the intermediate values, have been chosen for the parameter $u$. In the tables, $q$ is the cosmic acceleration, $z$ is cosmological redshift, $R$ is the scalar curvature of space–time, $t - t_{st}$ is the proper time, and $H$ is the Hubble parameter.

**TABLE I.** Space kinematics at the maximum energy density $\rho_{gr\max}$ = 5.2 × $10^{109}$ J·$m^{-3}$.

| $T\sqrt{\gamma_{\min}}$= $t_{pl}$ s; $\rho_{gr\max}$ = 5.2 × $10^{109}$ J·$m^{-3}$; $\sigma$ = 0.2501278984; $u^0$ = 6.118625359 | | | | | |
|---|---|---|---|---|---|
| $u$ | $q$ | $z$ | $R$, $m^{-2}$ | $t - t_{st}$, s | $H$, $s^{-1}$ |
| 6.118625359 | 0.7599 | 0 | −5.589 × $10^{-52}$ | 4.358 × $10^{17}$ | 2.181 × $10^{-18}$ |
| 1.362294111 | 0 | 0.84987 | −6.308 × $10^{-52}$ | 1.876 × $10^{17}$ | 3.074 × $10^{-18}$ |
| 0.853462941 | −0.5 | 1.41598 | −6.144 × $10^{-52}$ | 1.129 × $10^{17}$ | 4.290 × $10^{-18}$ |
| 0.8 | −0.5819 | 1.52552 | −5.890 × $10^{-52}$ | 1.029 × $10^{17}$ | 4.594 × $10^{-18}$ |
| 0.7 | −0.7600 | 1.79266 | −4.732 × $10^{-52}$ | 8.275 × $10^{16}$ | 5.435 × $10^{-18}$ |
| 0.6 | −0.9789 | 2.20159 | −6.939 × $10^{-53}$ | 6.051 × $10^{16}$ | 7.019 × $10^{-18}$ |
| 0.5 | −1.2496 | 2.93915 | 1.977 × $10^{-51}$ | 3.650 × $10^{16}$ | 1.089 × $10^{-17}$ |
| 0.4 | −1.5775 | 4.83051 | 3.079 × $10^{-50}$ | 1.305 × $10^{16}$ | 2.826 × $10^{-17}$ |
| 0.35 | −1.7543 | 7.80386 | 3.648 × $10^{-49}$ | 4.161 × $10^{15}$ | 8.511 × $10^{-17}$ |
| 0.3 | −1.9156 | 24.3239 | 1.843 × $10^{-46}$ | 1.963 × $10^{14}$ | 1.737 × $10^{-15}$ |
| 0.28 | −1.9643 | 87.6127 | 3.105 × $10^{-43}$ | 4.848 × $10^{12}$ | 6.945 × $10^{-14}$ |
| 0.265826306 | −1.9880 | 1090 | 9.993 × $10^{-37}$ | 2.719 × $10^{9}$ | 1.230 × $10^{-10}$ |
| 0.250127898 | −1.9985 | 8.09106 × $10^{10}$ | 1.484 × $10^{11}$ | 7.062 × $10^{-15}$ | 4.723 × $10^{13}$ |
| 0.146537059 | −0.5 | 2.16717 × $10^{20}$ | −9.435 × $10^{66}$ | 3.785 × $10^{-43}$ | 5.317 × $10^{41}$ |
| 0.137705891 | 0 | 2.24648 × $10^{20}$ | −2.067 × $10^{67}$ | 3.125 × $10^{-43}$ | 5.565 × $10^{41}$ |
| 0.125063950 | 1 | 2.33685 × $10^{20}$ | −4.321 × $10^{67}$ | 2.426 × $10^{-43}$ | 5.689 × $10^{41}$ |
| 0 | ∞ | 2.58860 × $10^{20}$ | −4.788 × $10^{68}$ | 0 | 0 |

**TABLE II.** Space kinematics at the maximum energy density $\rho_{gr\max}$ = 2 × $10^{49}$ J·$m^{-3}$.

| $\rho_{gr\max}$ = 2 × $10^{49}$ J·$m^{-3}$; $T\sqrt{\gamma_{\min}}$= 8.6912868 × $10^{-14}$ s; $\sigma$ = 0.2505131772; $u^0$ = 6.116607675 | | | | | |
|---|---|---|---|---|---|
| $u$ | $q$ | $z$ | $R$, $m^{-2}$ | $t - t_{st}$, s | $H$, $s^{-1}$ |
| 6.116607675 | 0.75979 | 0 | −5.588 × $10^{-52}$ | 4.358 × $10^{17}$ | 2.181 × $10^{-18}$ |
| 1.362058100 | 0 | 0.84978 | −6.145 × $10^{-52}$ | 1.876 × $10^{17}$ | 3.074 × $10^{-18}$ |
| 0.853190333 | −0.5 | 1.41607 | −9.218 × $10^{-52}$ | 1.129 × $10^{17}$ | 4.291 × $10^{-18}$ |
| 0.8 | −0.58143 | 1.52504 | −5.894 × $10^{-52}$ | 1.030 × $10^{17}$ | 4.592 × $10^{-18}$ |
| 0.7 | −0.75942 | 1.79220 | −4.740 × $10^{-52}$ | 8.281 × $10^{16}$ | 5.433 × $10^{-18}$ |
| 0.6 | −0.97810 | 2.20058 | −7.194 × $10^{-53}$ | 6.057 × $10^{16}$ | 7.014 × $10^{-18}$ |
| 0.5 | −1.2485 | 2.93699 | 1.0963 × $10^{-51}$ | 3.656 × $10^{16}$ | 1.088 × $10^{-17}$ |
| 0.4 | −1.5757 | 4.82107 | 3.042 × $10^{-50}$ | 1.313 × $10^{16}$ | 2.813 × $10^{-17}$ |
| 0.35 | −1.7519 | 7.76490 | 3.545 × $10^{-49}$ | 4.218 × $10^{15}$ | 8.404 × $10^{-17}$ |
| 0.3 | −1.9124 | 23.6505 | 1.564 × $10^{-46}$ | 2.130 × $10^{14}$ | 1.602 × $10^{-15}$ |
| 0.28 | −1.9607 | 78.6976 | 1.638 × $10^{-43}$ | 6.670 × $10^{12}$ | 5.054 × $10^{-14}$ |
| 0.263724335 | −1.9863 | 1090 | 9.820 × $10^{-37}$ | 2.741 × $10^{9}$ | 1.221 × $10^{-10}$ |
| 0.250513177 | −1.9939 | 5.92654 × $10^{5}$ | 2.294 × $10^{-20}$ | 17.963 | 0.001859 |
| 0.146809667 | −0.5 | 1.84758 × $10^{10}$ | −3.638 × $10^{6}$ | 6.102 × $10^{-13}$ | 3.301 × $10^{11}$ |
| 0.137941901 | 0 | 1.91536 × $10^{10}$ | −7.974 × $10^{6}$ | 5.037 × $10^{-13}$ | 3.456 × $10^{11}$ |
| 0.125256589 | 1 | 1.99255 × $10^{10}$ | −1.6669 × $10^{7}$ | 3.910 × $10^{-13}$ | 3.533 × $10^{11}$ |
| 0 | ∞ | 2.20739 × $10^{10}$ | −1.8450 × $10^{8}$ | 0 | 0 |

Thus, instead of the standard cosmological model (SCM), in this case, we have a continuum of cosmological models parameterized by the value of the maximum energy density $\rho_{gr\max}$. Comparison of the data in Tables I and II shows that the results of the calculation are in good agreement, at least up to redshift of the last-scattering surface,

$$z(0.2647 \pm 0.0011) = 1090, \quad z(u) = \left(\sqrt{\frac{\gamma(u^0)}{\gamma(u)}}\right)^{1/3} - 1, \tag{3.43}$$

despite a difference in the value of the maximum energy density of more than sixty orders. This circumstance excludes doubts about the possibility of an unambiguous description of the evolution of space in this range of redshift variation. It should be noted that the “last scattering” occurred less than 100 years after the beginning of the evolution process, as opposed to 373000 years in the ΛCDM model (Ref. 18, pp. 110, 111).

Significant differences between the models exist only at large values of $z$. The scalar curvature has a definite final value at the moment of the beginning of evolution; therefore, it is possible to determine the characteristic initial size as the reciprocal of the root of the modulus of curvature. This size depends on the value $\rho_{gr\max}$, and for the energy ranges considered in Tables I and II, can be from $10^{-34}$ to $10^{-4}$ meters.

### E. Geodesics, the problem of singularities and entropy of a homogeneous space

The lines $x^1 = x^2 = x^3 =$ const are geodesics for metric (3.27), as for the Friedmann–Lemaître–Robertson–Walker metric, and in each point, it is possible to introduce the concomitant coordinate system where *the variable t defined above will be a proper time*.

Substituting the Christoffel symbols (3.2) for metric (3.27) in the geodesic equations $x^\mu(\xi)$ with the natural parameter $\xi$,

$$\frac{d^2x^\mu}{d\xi^2} + \Gamma^\mu_{\nu\lambda}\frac{dx^\nu}{d\xi}\frac{dx^\lambda}{d\xi} = 0,$$

and integrating the derived equations, we find:

$$\frac{dx^m}{d\xi} = A^m\gamma^{-\frac{1}{3}}(x^0),\ \sqrt{g_{00}(x^0)}\frac{dx^0}{d\xi} = \pm\sqrt{A^2\gamma^{-\frac{1}{3}}(x^0) + B},\ A^2 = A^m\delta_{mn}A^n, A^m, B = \text{const.} \tag{3.44}$$

The edge points defined by the equation $t = t_{st}$ are the boundary of the found space–time manifold. On the boundary, $u(t_{st}) = 0$ and the cosmic acceleration (3.29), which is an invariant observable quantity, becomes infinite. In this regard, any geodesic extending to the boundary, with a finite value of $\xi$, will be confronted with an unremovable singularity. Consequently, *the found manifold is maximally extendable along geodesics up to the boundary*.

The authors (Ref. 19, p. 3) discussed the possibility of such a situation at the boundary of space-time: “One can think of a singularity as a place where our present laws of physics break down. Alternatively, one can think of it as representing part of the edge of space-time, but a part which is at a finite distance instead of at infinity. On this view, singularities are not so bad, but one still has the problem of the boundary conditions. In other words, one does not know what will come out of the singularity.” This issue is considered from the perspective of the quantum theory of gravity in Section V.

As for the manifold itself, in GR in many situations, it is sufficient to satisfy certain inequalities for the energy–momentum tensor in order to prove its singularity regardless of its specific form (Ref. 19, 4.3). It is customary to impose energy conditions of two types on the components of the energy–momentum tensor.

The weak energy condition will hold if $\rho \geq 0$, $\rho + p_m \geq 0$ ($m = 1, 2, 3$). “These inequalities are very reasonable requirements and are satisfied by all experimentally detected fields.” (Ref. 19, 4.3)

The dominant energy condition will hold if $\rho \geq 0$, $\rho \geq p_m \geq -\rho$ ($m = 1, 2, 3$). “This holds for all known forms of matter and there is in fact good reason for believing that this should be the case in all situations.” (Ref. 19, 4.3)

In the proposed theory, the situation is diametrically opposite to GR. *It turns out that it is possible to choose such a restriction on the parameters of the gravitational field, at which only a boundary will be singular.*

Relations (3.32), (3.36) imply

$$\rho_{gr}(u) \geq 0,\ \rho_{gr}(u) - p_{gr}(u) = \frac{c^2}{48\pi G T^2 \gamma(u)}[(4u-1)^2 + 4\sigma - 1].$$

It was noted above that for σ > 1/4, the metric has no singularities. In this case, from the previous relation it follows $\rho_{gr} > p_{gr}$. Thus, under the condition

$$\rho_{gr} \geq 0, \quad \rho_{gr}(u) > p_{gr}(u)\,, \tag{3.45}$$

the constructed manifold does not have the singularity outside a boundary.

According to (3.44), the velocity 4-vector is defined in the concomitant coordinate system along the geodesic

$$v^{\mu} = \frac{dx^{\mu}}{ds} = (g_{00}{}^{-\frac{1}{2}},\ 0,\ 0,\ 0).$$

In this case, it follows from the adiabatic equation, in consideration of (2.4), (2.5), (3.17), and (3.21), that:

$$s_{gr} = a\frac{k}{l_{pl}^2} v^{\lambda}(x)\frac{\partial \ln\sqrt{-g}}{\partial x^{\lambda}} = \frac{a\times k}{2 l_{pl}^2 c T\sqrt{\gamma(u)}} = \frac{\sigma\times k}{2 l_{pl}^2 c T\sqrt{\gamma(u)}}\,. \tag{3.46}$$

In the last equality, we identified an unknown constant $a$ with the only in the theory dimensionless parameter $\sigma$ (3.18) that characterizes the found space–time manifold. It follows from (3.46) that the entropy density of the manifold currently depends rather weakly on the maximum energy density, and, at $\rho_{gr\max}$ = 2 × $10^{49}$ J·m$^{-3}$, is equal to

$$s_{gr}(u^0) = \frac{\sigma\times k}{2 l_{pl}^2 c T\sqrt{\gamma_{\min}} f(u^0)} \approx 1.7\times 10^{42}\ \mathrm{k}\cdot\mathrm{m}^{-3}.$$

This value is 18 orders of magnitude greater than the contribution of all remaining entropy sources considered within the framework of GR.[20]

In view of (3.44), for an observer resting at the origin of the coordinates and connected by a 0-geodesic ($B$ = 0) with a concomitant point, the distance is determined (as in GR) by the relation[7]

$$d(t) = c\times a(t^0)\times\int_t^{t^0}\frac{dt}{a(t)}, \tag{3.47}$$

where $a(t)$ is the scale factor and $t$ is the proper time. The factor $a(t) = \gamma^{1/6}(t)$ is determined in the case under consideration by the relations given above; after the discovery of cosmic acceleration, in GR, it is determined within the framework of the ΛCDM model.[7] The parameters of this model were selected proceeding from a condition of providing the best agreement with all sets of experimental data that are available at the present time. The numerical values of these parameters as of 2013 are given in Ref. 18 and further used in calculating the dependences shown in the figures.

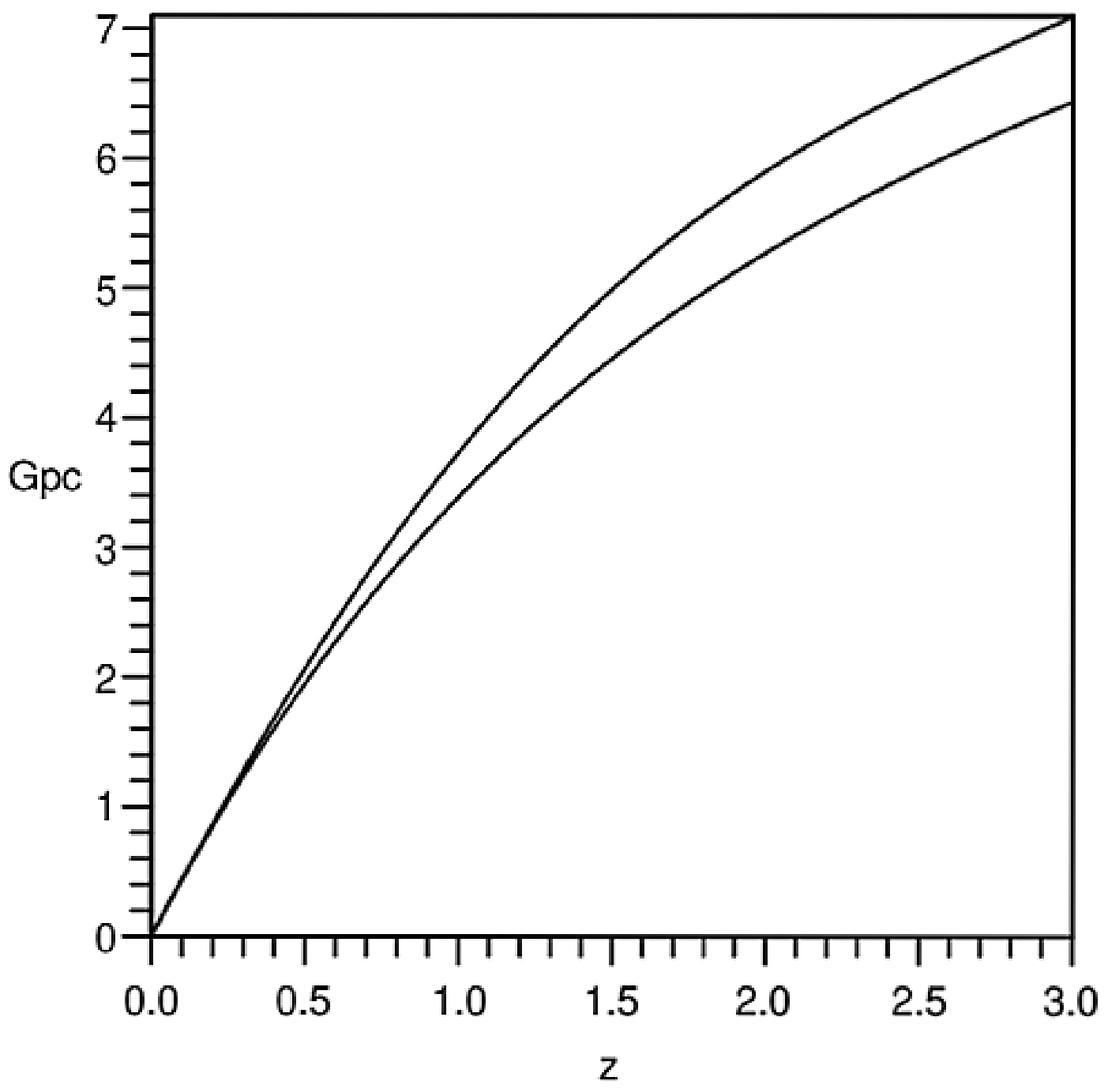

**FIG. 1.** Distance to an object (Gpc) depending on its redshift, as calculated by (3.47) for this theory and the ΛCDM models. The upper curve displays this theory, the lower curve displays the ΛCDM model.

All data sets relating to the dependence of distance on redshift that were available at the time of writing were given in a graphical form (Ref. 18, p. 364, Fig. 26.1). Comparison with these data sets shows that both dependences presented in Fig. 1 lie in the range of error of the experimental data. Moreover, this error grows faster with increasing *z* than the deviation between the calculated curves. When the above dependence is continued to the region of large values of *z*, its course will be defined by the maximum energy density of the gravitational field, which is unknown at the present time. Fig. 2 shows the relative distance $D_{rel}$ calculated as the ratio of the quantities presented in Fig. 1 over a wider range of redshift. The relatively small value of the deviation is associated with the integral nature of the dependence of the distance on redshift. For a local parameter, such as the Hubble parameter, the situation is different (Fig. 3). In this case, as the comparison of the calculation results with the experimental data shows (Fig. 4; see Ref. 21, p. 20), both dependences are also within the limits of the experimental error for $z < 2.5$. Figure 4 shows the quantity $H_{rel}=H_{\lambda cdm}/H$ equal to the ratio of the quantities presented in Fig. 3 over a wider range of redshift. The discrepancy between them increases dramatically at large redshifts, as shown in Fig. 4. Thus, only one of the two theories can be valid.

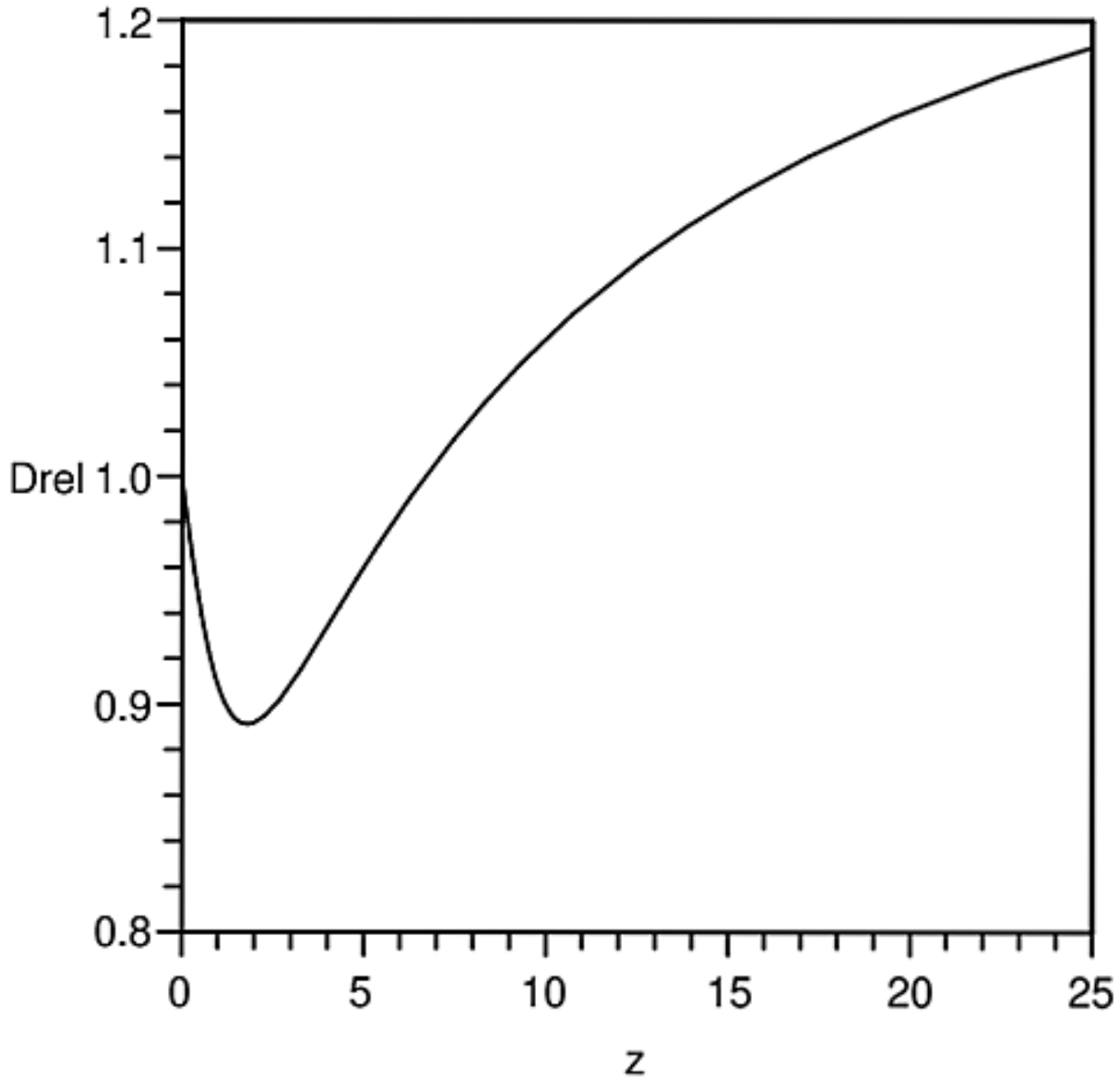

**FIG. 2.** Deviation of the ratio of the distances from unity calculated according to GR and this theory (vertically) depending on the value of redshift.

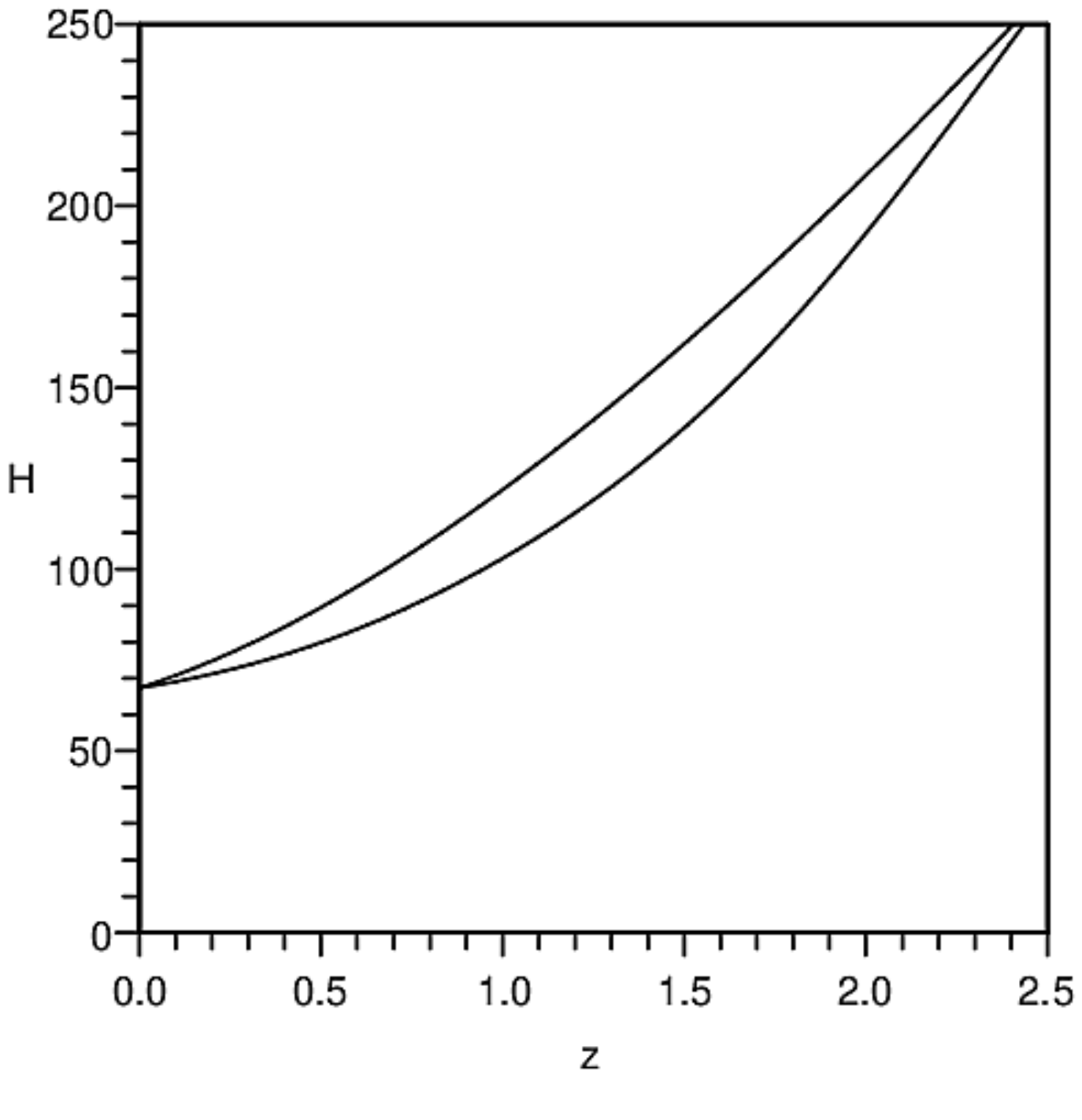

**FIG. 3.** Dependence of the Hubble parameter (km·s$^{-1}$·Mpc$^{-1}$) on redshift. The upper curve represents the ΛCDM model, the lower curve represents this theory.

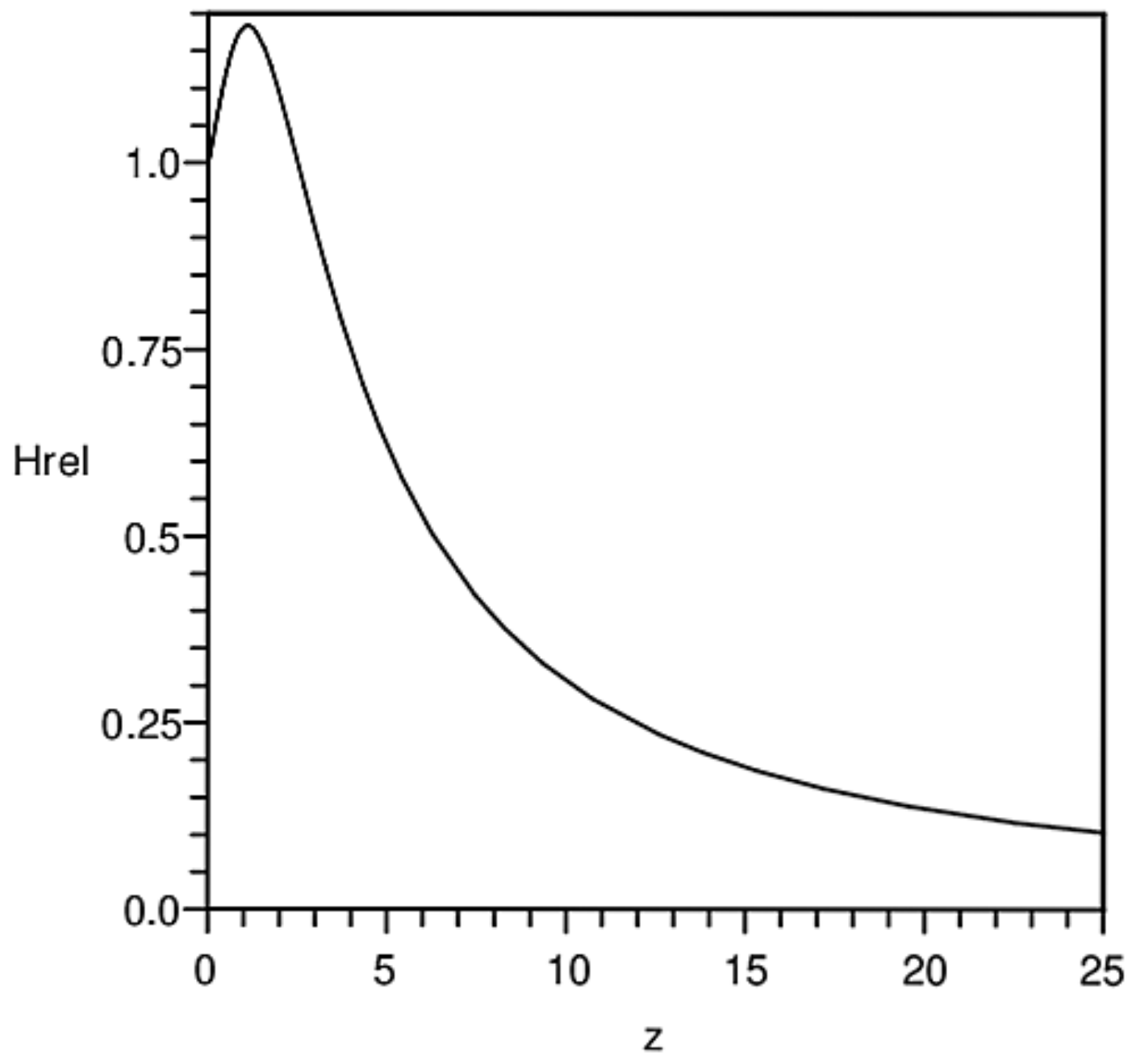

**FIG. 4.** Dependence of the ratio of the Hubble parameter in the ΛCDM model to its value in this theory.

It is essential that the dependence does not have free parameters in this region of redshifts, is determined only by the initial values at $z = 0$, and, as can be seen from Table I, is valid up to the initial instant of time. The hypotheses about the existence of dark energy and dark matter are introduced in GR within the framework of the ΛCDM model to ensure agreement with experiments. The hypothesis about the existence of inflatons is introduced to describe the dependence in the region of large values of $z$.

## F. Temperature of the homogeneous gravitational field

The gravitational field in empty space possesses the characteristics inherent in a material medium: energy, pressure, and entropy. By virtue of the general laws of thermodynamics, another characteristic of the state of a medium is temperature, the change of which in an equilibrium process without a heat supply is associated with a change of the pressure by the relation[22]

$$s_{gr}\frac{d\theta_{gr}}{dt} = \frac{dp_{gr}}{dt}. \quad (3.48)$$

This equation allows us to determine the temperature of empty space from the found dependences of the pressure (3.36) and entropy density of the gravitational field (3.46) on $u$. Substituting the corresponding relationships into (3.48), we have

$$-\frac{c^2}{48\pi G T^2}\sqrt{\gamma(u)}d\left(\frac{2u^2-2u+\sigma}{\gamma(u)}\right) = \frac{\sigma\times k}{2l_{pl}^2 cT}d\theta_{gr}. \quad (3.49)$$

Integrating this equation taking into account the dependence $\gamma(u)$ (3.21), we find

$$\theta_{gr}(u) = \frac{\hbar}{12\pi\times k\times T\sqrt{\gamma_{\min}}}\int_0^u \frac{1}{\sigma f(u)}\times\frac{\sigma-2u(1-\sigma)}{\sigma-2u+4u^2}du. \quad (3.50)$$

The value of the integral for the data in Tables I and II is the same and is 0.5 (with an accuracy of eight significant figures) at $u = u^0$ (the present moment of time). At the same time, the temperature of the gravitational field $\theta_{gr}(u^0)$ changes from 1.879 × $10^{30}$ K in the first case to 1.166 K in the second. Fig. 5 implicitly shows the dependence of the relative temperature $T_{rel}= \theta_{gr}(u)/\,\theta_{gr}(u^0)$ on time.

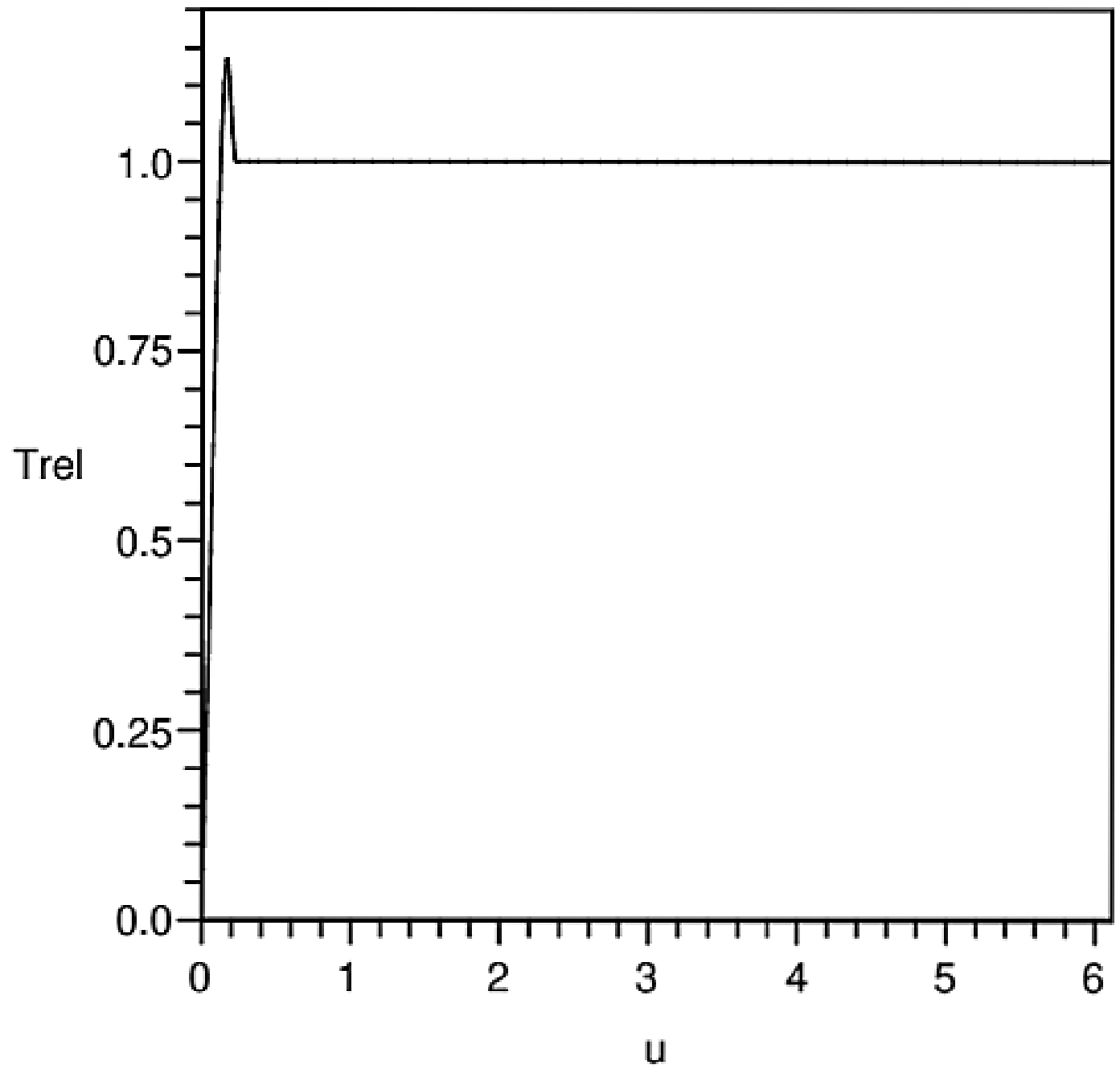

**FIG. 5.** Ratio of the current temperature to its value at the present time as a function of *u* (3.19).

The relationship between the dimensionless rate of change of the volume factor $u$ and the proper time is seen in Tables I and II. It should be noted that for both variants presented in the table, the calculated dependences are almost identical. Aside from a short initial time interval, the temperature of the manifold remains at a constant level equal to its current value; that is, the gravitational field has had a constant temperature for almost 14 billion years up to the present moment, and it will continue to have this value until achieving complete equilibrium. It acts as a thermostat for the Universe.

The temperature of such a thermostat can be estimated by observing the temperature change of the bodies in thermal contact with it. Their temperature should asymptotically tend to the temperature of the thermostat when approaching the equilibrium state. In particular, if we consider the temperature of the cosmic microwave background (CMB) radiation, under adiabatic cooling it will tend to the temperature of the thermostat over time, but not to zero. At present, the temperature of the CMB radiation is $\theta_{rel}^0$ = 2.7255 K, and no changes have been recorded in the course of its adiabatic cooling. From this, we can conclude that the temperature of the gravitational field $\theta_{gr}(u^0)$ is less than $\theta_{rel}^0$.

Equation (3.50) shows that

$$T\sqrt{\gamma_{\min}} = \frac{0.5\times\hbar}{12\pi k\theta_{gr}(u^0)}\,. \qquad (3.51)$$

Using this relation and (3.41), it is possible to relate the maximum value of the global energy density $\rho_{gr\max}$ to the present temperature value:

$$\rho_{gr\max} = \frac{3\pi}{2e^2}\times\frac{c^2}{G}\times\left(\frac{k\theta_{gr}(u^0)}{\hbar}\right)^2. \qquad (3.52)$$

If we take the temperature $\theta_{gr}(u^0)$ as being equal to the CMB radiation temperature at the present time $\theta_{rel}^0$= 2.7255 K for estimation, then $\rho_{gr\max} < 1 \times 10^{50}$ J·m$^{-3}$ ~ (1.5 TeV)$^4$. This is close to the value in the variant of evolution presented in Table II and *differs strikingly from the SCM, in which the energy density can reach a value 64 orders of magnitude greater* ($10^{19}$ GeV)$^4$. Perhaps this is exactly the reason for the absence in the Universe of the hypothetical forms of matter that are not found in experiments at the Large Hadron Collider (LHC).

## IV. BASIC MODEL OF THE EVOLUTION OF A HOMOGENEOUS AND ISOTROPIC UNIVERSE

An increase in the intensity of the gravitational field during the evolution process will inevitably lead to the appearance of new structures of matter. Let us consider phenomenologically the influence of matter on the process of the evolution of the Universe.

In the presence of matter fields, the general action, along with $S_{gr}$, must include the action of matter $S_{mat}$.

$$S = S_{gr} + S_{mat}.$$

It was proved in GR[15] that the scalar curvature must enter into action with a minus sign, however this proof does not hold in the presence of the constraint and the condition $g(x) \neq$ const.
Sec. V shows that the action $S_{gr}$ can only have a minimum when the plus sign is chosen.

$$S_{gr} = \frac{c^3}{16\pi G}\int (R+Q)\sqrt{-g}d^4x,\; Q = \frac{1}{\sqrt{-g}}\frac{\partial\sqrt{-g}}{\partial x^\mu}g^{\mu\nu}\frac{\partial \Phi}{\partial x^\nu}.$$

As shown in Sec. III, there is a unique homogeneous space–time with an isotropic metric of the form

$$ds^2 = g_{00}(x^0)(dx^0)^2 - \gamma^{1/3}(x^0)dx^m dx^n \delta_{mn}.$$

Let matter be born at some moment in this space–time. Owing to its homogeneity and isotropy, we write the tensor of the average energy–momentum density of matter in the form $(\varepsilon_{mat})_\mu^\nu = \text{diag}(\rho_{mat,}\ -p_{mat,}\ -p_{mat,}\ -p_{mat})$. According to Eqs. (2.2), this tensor should be included in Eqs. (3.7)–(3.9).

In the presence of matter, taking into account the change in the sign of the action, the gravitational field equations (3.7)–(3.9) will take the forms

$$\frac{d}{dx^0}\left(\frac{1}{g_{00}}\frac{d\sqrt{\gamma g_{00}}}{dx^0}\right) = 0,$$

$$\frac{1}{\sqrt{g_{00}}}\frac{d}{dx^0}\left(\frac{1}{\gamma\sqrt{g_{00}}}\frac{d\gamma}{dx^0}\right) + \frac{1}{6g_{00}}\left(\frac{1}{\gamma}\frac{d\gamma}{dx^0}\right)^2 = -\sqrt{\gamma g_{00}}\frac{d}{dx^0}\left(\frac{1}{g_{00}\sqrt{\gamma g_{00}}}\frac{d\Phi}{dx^0}\right) + \frac{8\pi G}{c^4}(\rho+3p)_{mat},$$

$$\frac{1}{\sqrt{\gamma g_{00}}}\frac{d}{dx^0}\left(\sqrt{\frac{\gamma}{g_{00}}}\frac{1}{3\gamma}\frac{d\gamma}{dx^0}\right)\delta_k^p = -\delta_k^p\frac{1}{\sqrt{\gamma g_{00}}}\frac{d}{dx^0}\left(\sqrt{\frac{\gamma}{g_{00}}}\frac{d\Phi}{dx^0}\right) - \frac{8\pi G}{c^4}(\rho-p)_{mat}\delta_k^p.$$

Repeating all the computations taking into account these additional terms, instead of (3.20), we obtain the integro-differential equation

$$8\gamma u\frac{du}{d\gamma} = 4u^2 - 2u + \sigma + M(u,\gamma,\frac{d\gamma}{du}), \tag{4.1}$$

where

$$M(u,\gamma,\frac{d\gamma}{du}) = \frac{48\pi G T^2}{c^2}\left(\gamma(\rho+p)_{mat} - \frac{1}{4}\int_0^u (\rho-p)_{mat}\left(\frac{d\gamma}{du}\right)\frac{du}{u}\right),$$

and it is supposed that *the pressure and density of matter are equal to zero at the initial time*.

The equations for cosmic acceleration, energy density, pressure, and scalar curvature of space are also modified in this case; instead of (3.29), (3.32), (3.36), and (3.39), we have

$$q = 1 - \frac{3}{2u} + \frac{3\sigma}{4u^2} + \frac{3}{4u^2}M(u,\gamma,\frac{d\gamma}{du}), \tag{4.2}$$

$$\rho_{gr} - \rho_{mat} = \frac{c^2}{24\pi G T^2}\frac{u^2}{\gamma(u)} = \frac{3c^2 H^2(u)}{8\pi G} \equiv \rho_{cr}(u), \tag{4.3}$$

$$p_{gr} - p_{mat} = -\frac{c^2}{48\pi G T^2}\frac{1}{\gamma(u)}\left[2u^2 - 2u + \sigma + M(u,\gamma,\frac{d\gamma}{du})\right], \tag{4.4}$$

$$R = -\frac{1}{2c^2T^2\gamma(u)}\left[\frac{8}{3}u^2 - 2u + \sigma + M(u,\gamma,\frac{d\gamma}{du})\right]. \tag{4.5}$$

According to observation data, the Universe currently contains macroscopic matter, electromagnetic radiation, and neutrinos. These components interact weakly both with each other and with the gravitational field. In this case, the conservation laws for each type of matter are satisfied separately, therefore the covariant derivative of the tensor $(\varepsilon_{mat})_\mu^\nu$ must be equal to zero.

$$\frac{1}{\sqrt{-g}}\frac{\partial}{\partial x^\nu}(\sqrt{-g}(\varepsilon_{mat})_\mu^\nu) - \Gamma_{\rho\mu}^\lambda(\varepsilon_{mat})_\lambda^\rho = 0.$$

Substituting into this equation the expressions for connectivity (3.2) in the case of an isotropic metric, we obtain

$$\frac{d\rho_{mat}}{dx^0} = -(\rho + p)_{mat}\frac{1}{\sqrt{\gamma}}\frac{d\sqrt{\gamma}}{dx^0}$$

The pressure can be considered equal to zero for baryonic matter, $p = \rho/3$ for electromagnetic radiation, and for neutrinos, a similar relation will be valid as long as it is possible to neglect their mass. Under these conditions, for the components of matter, we obtain

$$\rho_b = \rho_b^0\frac{\sqrt{\gamma^0}}{\sqrt{\gamma}},\ \rho_\gamma = \rho_\gamma^0\left(\frac{\sqrt{\gamma^0}}{\sqrt{\gamma}}\right)^{4/3},\ \rho_\nu = \rho_\nu^0\left(\frac{\sqrt{\gamma^0}}{\sqrt{\gamma}}\right)^{4/3}. \tag{4.6}$$

The values relating to the current time are denoted by the superscripts.

It is known that the energy densities of the two first components are respectively equal to $\Omega_b = 0.0499$ and $\Omega_\gamma = 5.46 \times 10^{-5}$ of the critical energy density at the present time (Ref. 18, pp. 110, 111). The data are less well defined for neutrinos, and $\Omega_\nu < 5.52 \times 10^{-3}$. Then, to estimate the maximum degree of the influence of matter on the evolution process, exactly this value of the relative density of neutrinos will be used.

At times not too far from the present, we have the following dependence of the average energy density and pressure of matter on the volume factor:

$$\rho_{mat} = \rho_{cr}^0\left[\Omega_b\frac{\sqrt{\gamma^0}}{\sqrt{\gamma}} + \ \Omega\left(\frac{\sqrt{\gamma^0}}{\sqrt{\gamma}}\right)^{4/3}\right],\ p_{mat} = \frac{\rho_{cr}^0}{3}\ \Omega\left(\frac{\sqrt{\gamma^0}}{\sqrt{\gamma}}\right)^{4/3},\ \Omega = \Omega_\gamma + \Omega_\nu. \tag{4.7}$$

The functional on the right-hand side (4.2–4.6) describes the inverse effect of matter on the metric. The functional is equal to zero at the beginning of evolution, and all the energy is concentrated in the gravitational field; therefore, to a first approximation, the inverse action can be neglected. Suppose that

$$M^{(1)}(u,\gamma,\frac{d\gamma}{du}) = 0,$$

where the index in parentheses indicates the approximation number. In this case, the change of the volume factor and its derivative will continue to be described by relations (3.20) and (3.21), and the critical density by relation (4.3). Thus, in this approximation, the energy density and matter pressure can be considered known functions of $u$ at $u_b \le u$:

$$\rho_{mat}(u) = \rho_{cr}^0\frac{f(u^0)}{f(u)}\left[\Omega_b + \ \Omega\left(\frac{f(u^0)}{f(u)}\right)^{1/3}\right],\ p_{mat} = \frac{\rho_{cr}^0}{3}\ \Omega\left(\frac{f(u^0)}{f(u)}\right)^{4/3},\ \Omega = \Omega_\gamma + \Omega_\nu. \tag{4.8}$$

## A. Energy density of matter in the very early Universe

The conditions under which relations (4.8) are valid are violated at $0 \le u \le u_b$ (the very early Universe). The reason for this is the extremely high energy density of the gravitational field, reaching the level of the energy density of the LHC (as shown in Sec. III). Under these conditions, in addition to the particles listed, other components of the standard model of elementary particles and fields will also be born.

With this in mind, we redefine the dependences $\rho_{mat}(u)$ at the beginning of the evolution process as follows. Since there are no other sources of energy than gravity during this period, we assume that it is proportional to $\rho_{gr}(u)$ with a dimensionless coefficient that depends on the energy density of the gravitational field:

$$\rho_{mat}(u) = \lambda \times \left(\frac{u}{f(u)}\right)^n \times \rho_{gr}(u),\ \ \lambda < 1,\ \ 0 \leq u \leq u_b,\ n \geq 0. \tag{4.9}$$

Excluding the gravitational energy density from relations (4.3) and (4.9), we have

$$\rho_{mat}(u) = \frac{\lambda \times u^n}{f^n(u) - \lambda \times u^n} \times \rho_{cr}(u),\ 0 \leq u \leq u_b,\ n \geq 0. \tag{4.10}$$

The constant $\lambda$ and the quantity $u_b$ are determined from the smooth conjugation conditions of dependences (4.8) and (4.10) at $u = u_b$. Equating separately the energy densities and their derivatives at $u = u_b$, we derive a system of two equations for determining $u_b$ and $\lambda$:

$$\frac{\lambda \times u_b^n}{f^n(u_b) - \lambda \times u_b^n} = \left(\frac{u^0}{u_b}\right)^2 \left[\Omega_b \frac{f(u_b)}{f(u^0)} + \Omega \left(\frac{f(u_b)}{f(u^0)}\right)^{2/3}\right], \tag{4.11}$$

$$4u_0^2 \times \left[\Omega_b \frac{f(u_b)}{f(u^0)} + \frac{4}{3}\Omega \left(\frac{f(u_b)}{f(u^0)}\right)^{2/3}\right] = \left(n + 2 + n\frac{\lambda \times u_b^n}{f^n(u_b) - \lambda \times u_b^n}\right) \frac{\lambda \times u_b^n (2u_b - \sigma)}{f^n(u_b) - \lambda \times u_b^n}\cdot \tag{4.12}$$

This system of equations has two different solutions, and the solution with the smaller value of $u_b$ is physically sensible. For this solution, $u_b \ll f(u_b) \ll f(u^0)$; therefore, Eq. (4.12) is simplified, and the solution takes the form

$$u_b \approx \frac{3(2+n) - \sqrt{9(2+n)^2 - 48(2+n) \times \sigma}}{16}.$$

The quantity $u_b$ determines that moment of world time when matter is being separated from the gravitational field. This is due to a decrease in the absolute value of the scalar curvature over time. This happens specifically at the time of its first conversion to zero. It follows from (3.39) that $R(u) = 0$ when

$$u = \frac{6 - \sqrt{36 - 96 \times \sigma}}{16}.$$

This expression is the same as $u_b$ at $n = 0$. From (4.11), we find the ratio of the energy density of matter to the energy density of the gravitational field at the time of its separation from matter for this value:

$$\lambda \approx \Omega \times \left(\frac{u^0}{u_b}\right)^2 \left(\frac{f(u_b)}{f(u^0)}\right)^{2/3},\ \ u_b \approx \frac{6 - \sqrt{36 - 96 \times \sigma}}{16}. \tag{4.13}$$

By substituting the values corresponding to the data of Table II into this relation, we find that this fraction was $2.754 \times 10^{-20}$. The energy density of the Universe is $1.390 \times 10^{49}$ J·m$^{-3}$ for $u = u_b$ at the time $t - t_{st} = 8.183 \times 10^{-13}$ s. When approaching the initial instant of time, the average energy density of matter decreases in accordance with (4.10), tending to zero together with the critical density.

As for the pressure of matter in the very early Universe, the contribution of baryonic matter to the total energy density is negligible during this period of time. The rest of the components of matter are ultrarelativistic, and therefore their pressure is one–third of the total energy density of matter (4.10). Hence it follows that in the very early Universe, as well as in later periods of time, both energy conditions of GR are satisfied for matter. The fundamental difference from GR is that not matter is the source of the gravitational field, but on the contrary, the intense gravitational field generates matter.

## B. Temperature history of the early and very early Universe

Let us now turn to the temperature history of the early Universe. In so doing, we will proceed from the main points set forth in Ref. 7. The early period includes the period of time when the temperature of electromagnetic radiation was in the range from $10^{11}$ K to 4000 K. It is stated[7] that the following relations between the density of entropy $s$, temperature $\theta$, and the scale factor $a$ for electromagnetic radiation and neutrino matter are valid (the corresponding quantities are marked with the subscript $\gamma$ or $\nu$):

$$s_\gamma(\theta)a^3 = \text{ const}, \; s_\gamma(\theta) = \frac{4}{3}a_B\theta_\gamma^3, \; s_\nu(\theta)a^3 = \text{ const}, \; s_\nu(\theta) = \frac{7}{2}a_B\theta_\nu^3 \,, \; a_B = \frac{\pi^2 k^4}{15\hbar^3 c^3}.$$

Using the expressions for the volume factor obtained in the previous section, we rewrite these relations in the form

$$s_\gamma(\theta)\gamma^{1/2}(u) = s_\gamma(\theta_b)\gamma^{1/2}(u_b); \; s_\gamma(\theta) = \frac{4}{3}a_B\theta_{\gamma b}^3\frac{\gamma^{1/2}(u_b)}{\gamma^{1/2}(u)} = \frac{4}{3}a_B\theta_{\gamma b}^3\frac{f(u_b)}{f(u)};$$

$$\theta_{\gamma b} = \theta_\gamma(u_b). \tag{4.14}$$

$$s_\nu(\theta)\gamma^{1/2}(u) = s_\nu(\theta_b)\gamma^{1/2}(u_b); \; s_\nu(\theta) = \frac{7}{2}a_B\theta_{\nu b}^3\frac{\gamma^{1/2}(u_b)}{\gamma^{1/2}(u)} = \frac{7}{2}a_B\theta_{\nu b}^3\frac{f(u_b)}{f(u)};$$

$$\theta_{\nu b} = \theta_\nu(u_b). \tag{4.15}$$

It is shown in Ref. 7 that Eq. (3.48) is also applicable to describing the change in the temperature of matter in the Universe. In this case, by virtue of the additivity of the contributions of components to the pressure and entropy density, Eq. (3.48) takes the form:

$$dp_{mat} = \big(s_\gamma(\theta) + s_\nu(\theta)\big)d\theta. \tag{4.16}$$

We apply this equation to the description of the initial stage of evolution at $0 \le u \le u_b \ll u^0$. In this case, as follows from (4.8), the contribution of baryons to the total energy density of matter is negligible, matter can be considered an ultrarelativistic medium, and taking into account (4.3) and (4.10), we can write the pressure of matter in the form

$$p_{mat}(u) = \frac{\lambda}{3}\times\frac{c^2}{24\pi G T^2\gamma_{\min}}\times\left(\frac{u}{f(u)}\right)^2, \; 0 \le u \le u_b.$$

Neglecting the possible differences in temperature of the components of matter, we write Eq. (4.16), which determines the change in temperature, as

$$\frac{\lambda}{3}\times\frac{c^2}{24\pi G T^2}\times\frac{f(u)}{f(u_b)}d\left(\frac{u}{f(u)}\right)^2 = \frac{29}{6}a_B\theta_b^3 d\theta. \tag{4.17}$$

Integrating this equation, taking into account definition (3.21) of the function $f(u)$ and its derivative, for we get

$$\theta^3(u_b)\times\theta(u) = \frac{\lambda}{29a_B}\times\frac{c^2}{6\pi G T^2\gamma_{\min}}\times\frac{1}{f(u_b)}\int_0^u\frac{(\sigma-2u)udu}{(4u^2-2u+\sigma)f(u)}. \tag{4.18}$$

Substituting (4.13) for the parameter $\lambda$ into this equation, we find the temperature $\theta(u_b)$ at the moment of separation of matter from the gravitational field.

$$\theta(u_b) = \left[\frac{1}{29a_B}\times\frac{c^2}{6\pi G T^2\gamma_{\min}}\times\frac{\Omega}{u_b}\left(\frac{u^{03}\times f(u_b)}{u_b^3\times f(u^0)}\right)^{2/3} I(u_b)\right]^{1/4}, \tag{4.19}$$

where

$$I(u_b) = \int_0^{u_b}\frac{(\sigma-2u)udu}{(4u^2-2u+\sigma)f(u)}.$$

The calculation for the values of the parameters corresponding to the data in Tables I and II gives $I(u_b) = 2.302(\pm 0.0005)\times 10^{-4}$. At the same time, $\theta(u_b) = 1.345\times 10^{21}$ K in the first case, and $\theta(u_b) = 1.145\times 10^{11}$ K in the second case.

According to the calculations given in Ref. 7, the radiation and neutrino temperatures coincide at $\theta = 10^{11}$ K. Below this temperature, neutrinos lose equilibrium with other particles and their temperature decreases, asymptotically tending to the value

$$\theta_\nu = \left(\frac{4}{11}\right)^{1/3} \theta_\gamma.$$

At the same time, the radiation temperature will decrease at $u > u_b$ with the growth of the scale factor according to the law[7]

$$\frac{\theta_\gamma(u)}{\theta_\gamma(u_b)} = \frac{\gamma^{1/6}(u_b)}{\gamma^{1/6}(u)} = \frac{f^{1/3}(u_b)}{f^{1/3}(u)}. \qquad (4.20)$$

Substituting expression (4.19) here, we find the dependence of the current value of radiation temperature on $u(t)$ at $u > u_b$:

$$\theta_\gamma(u) = \left[\frac{1}{29a_B} \times \frac{c^2}{8\pi G T^2 \gamma_{\min}} \times \frac{\Omega}{u_b}\left(\frac{u^{03}\cdot f(u_b)}{u_b^3\cdot f(u^0)}\right)^{2/3} \int_0^{u_b} \frac{(\sigma-2u)u^2 du}{(4u^2-2u+\sigma)f^2(u)}\right]^{1/4} \frac{f^{1/3}(u_b)}{f^{1/3}(u)}. \qquad (4.21)$$

### C. Relative density of neutrinos in the Universe

For $u = u^0$, that is, at the present moment in our Universe, the value of the temperature $\theta_\gamma(u^0)$ should be equal to the experimentally observed temperature of the CMB radiation, 2.7255 K. The calculation by (4.21) of the parameters values corresponding to the data in Table I gives $\theta_\gamma(u^0) = 6.6215$ K; for the data in Table II it gives $\theta_\gamma(u^0) = 6.6147$ K.

In the constructed continuum of models of the Universe for the two extreme cases, which differ in maximum energy density by 64 orders of magnitude, the temperatures of the CMB radiation practically coincide with each other at the same point of time (of our present), but they are more than double the value observed in our Universe. There are no free parameters in the described phenomenological model; therefore, such a discrepancy could mean its collapse, if not for one circumstance. As noted at the beginning of the section, the relative neutrino density according to the data in Ref. 18 (pp. 110, 111) is $\Omega_\nu < 5.52 \times 10^{-3}$, and it, unlike the radiation density $\Omega_\gamma = 5.46 \times 10^{-5}$, is not exactly determined, along with their sum $\Omega = \Omega_\gamma + \Omega_\nu$. This quantity is included in the expression for the temperature of the CMB radiation (4.21) in the form of a constant factor $\Omega^{1/4}$. Therefore, the discrepancy from experiment can be eliminated if a new value $\Omega^*$ is introduced instead of the old density value $\Omega = 5.575 \times 10^{-3}$:

$$\Omega^* = (\Omega_\nu^* + \Omega_\gamma) = 1.6068 \times 10^{-4},\ \Omega_\nu^* = 1.6068 \times 10^{-4} - 5.46 \times 10^{-5} = 1.0608 \times 10^{-4}. \qquad (4.22)$$

Thus, if the stated theory is correct, then *the currently unknown relative density of neutrinos in the Universe is equal to* $\Omega_\nu^* = 1.0608 \times 10^{-4}$. In this case, at $\rho_{gr\max} = 2 \times 10^{49}$ J·m$^{-3}$, $\theta_\gamma(u_b) = 4.719 \times 10^{10}$ K, and the maximum radiation temperature in the Universe $\theta_{\gamma\max} = 9.404 \times 10^{10}$ K is reached at $t - t_{st} = 4 \times 10^{-13}$ s. As shown in Sec. III, $\rho_{gr\max} < 2 \times 10^{50}$ J·m$^{-3}$. With this value of the maximum possible energy density, the temperature of matter in the Universe has never exceeded $\theta_{\gamma\max} = 1.230 \times 10^{11}$ K.

### D. Value of the relative density of matter observed in the Universe

Taking into account (4.8), the average relative energy density of matter (at the found value of the relative neutrino density $\Omega_\nu^*$ ) is

$$\text{quota}(u) = \frac{\rho_{mat}(u)}{\rho_{cr}(u)} = \frac{u^{02} f^2(u)}{u^2 f^2(u^0)}\left[\Omega_b \frac{f(u^0)}{f(u)} + \Omega^* \left(\frac{f(u^0)}{f(u)}\right)^{4/3}\right].$$

In view of (3.21) and (3.42), this quantity depends on time as shown in Fig. 6.

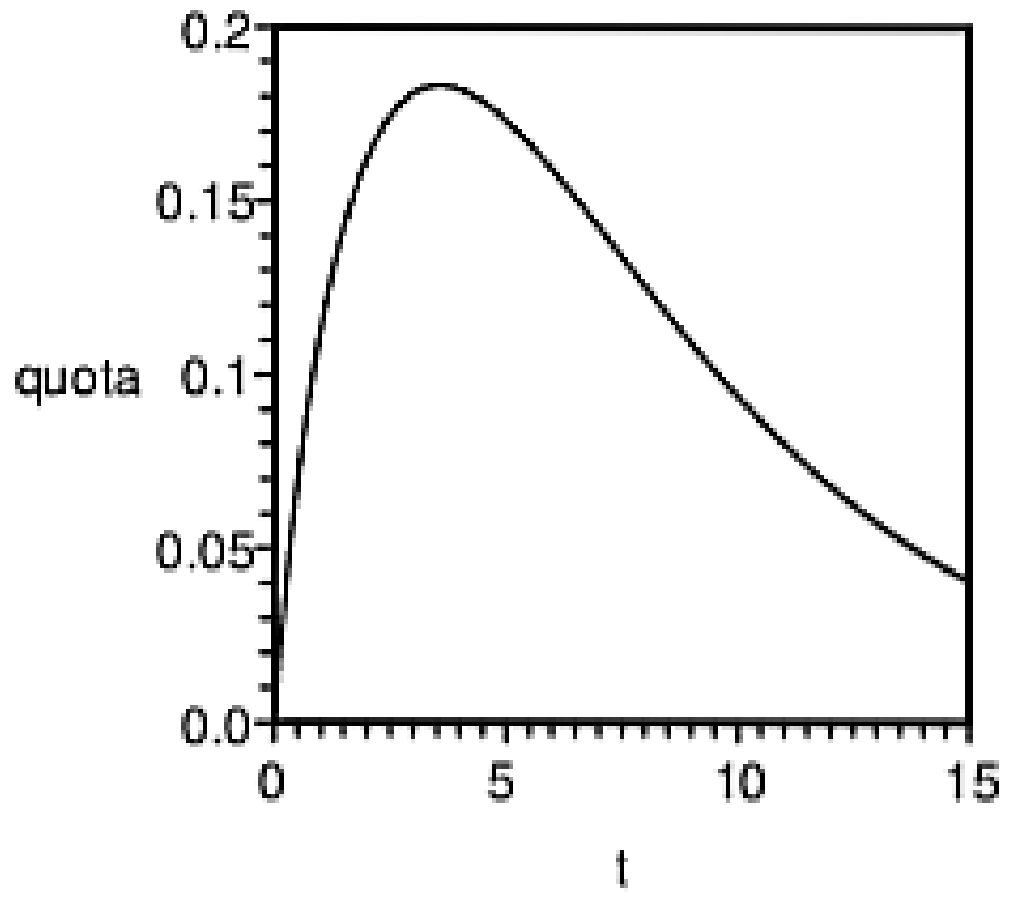

**FIG. 6.** Time dependence of the ratio of the average energy density of matter to the critical density in the Universe (in billions of years), $\rho_{gr\max} = 2 \times 10^{49}$ J·m$^{-3}$.

The maximum fraction of the energy of matter does not exceed 0.1832; at the present time, this value is less than 0.055 and continues to decrease with time. In contrast to GR, where the energy density of matter increases indefinitely with decreasing time, here it reaches its maximum and then begins to decrease.

The remainder and main part of the energy of the Universe is the energy of the gravitational field

$$\rho_{gr}(u) = \rho_{cr}(u)(1 + \text{quota}(u)).$$

In view of (3.44), this value can be related to redshift of observed objects. We consider two such objects located in a homogeneous and isotropic gravitational field, the energy density of which depends on their distance to an observer (Fig. 7). Let us mentally select the volume of a sphere with a radius equal to the distance between the objects. Such surrounding gravitational field does not affect the dynamics of these objects. However, the object on the surface will be under the influence of gravity of the mass of such sphere, consisting of the mass of the main object and the distributed mass of the gravitational field. It is this additional mass (energy), and not dark matter, that manifests itself in the character of the dependences of the rotation curves of gravitation-coupled objects.

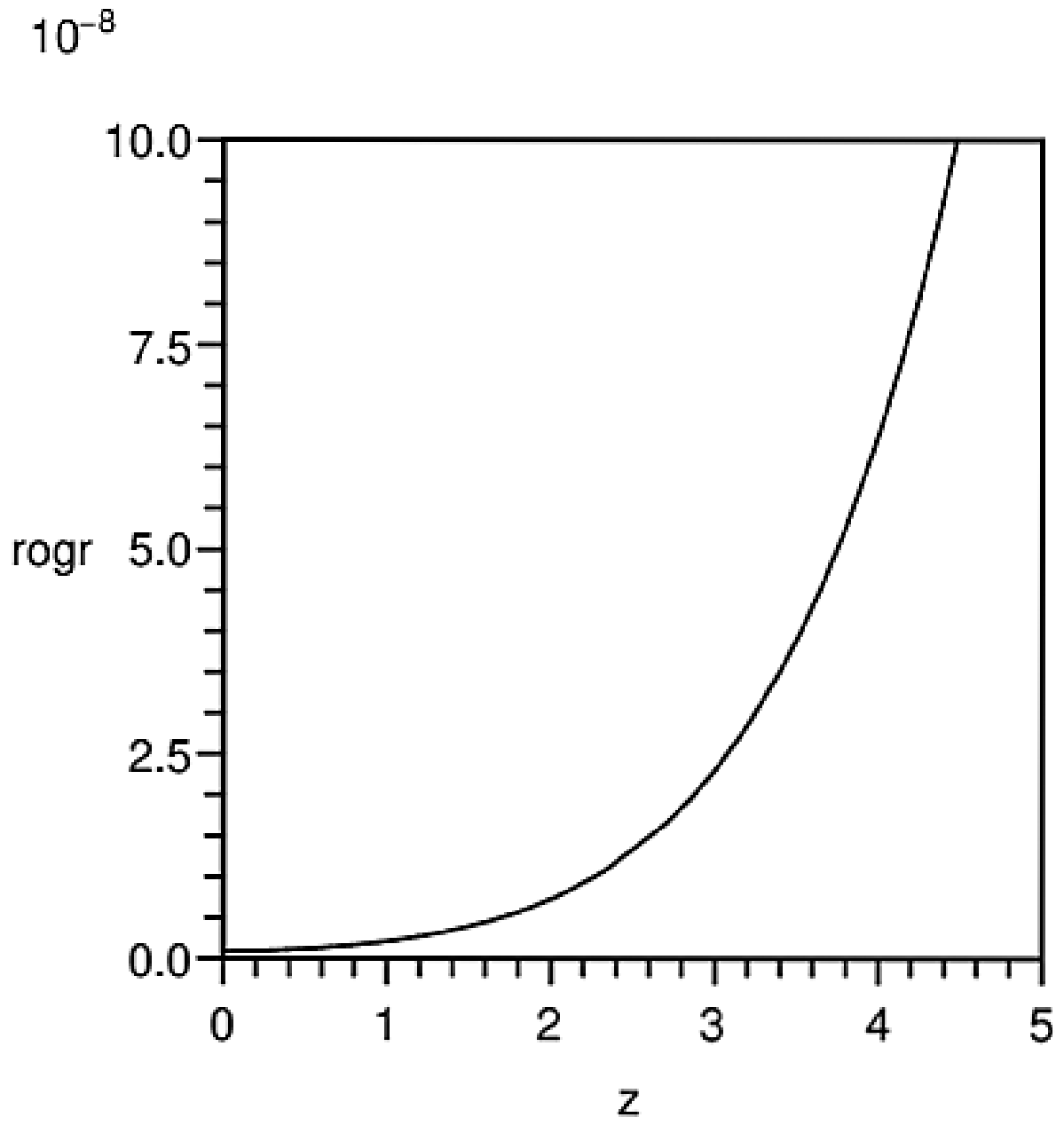

**FIG. 7.** Dependence of the energy density of the gravitational field $\rho_{gr}$ ($10^{-8}$J·m$^{-3}$) on redshift in the location of the observed gravitation-coupled objects, $\rho_{gr\max}$ = 2 × $10^{49}$ J·m$^{-3}$.

According to the data in Ref. 18 (pp.110, 111), the energy density of cold dark matter in GR is equal to

$$\rho_{cdm}(z) = \Omega_{cdm}\rho_{cr}^{0}(1+z)^3,\ \Omega_{cdm} = 0.265_{-0.17}^{+0.16}.$$

Fig. 8 shows the dependence of its ratio to the energy density of the gravitational field

$$\mathrm{beta}(z) = \frac{\rho_{cdm}(z)}{\rho_{gr}(z)}$$

over a wider range of redshift. As can be seen from Fig. 8, a reasonable agreement with experiment can be obtained in a certain region of redshifts when calculating the rotation curves of gravitationally bound objects using a hypothetical dark matter density. However, there is a vast region of these values where such a calculation will lead to erroneous results. There are no reasonable arguments for replacing the gravitational field, which is a real source of additional mass, with a hypothetical cold matter with a possible unpredictable error value. Thus, in contrast to GR, no new forms of matter besides those already known are required to describe the features of the evolution of the Universe.

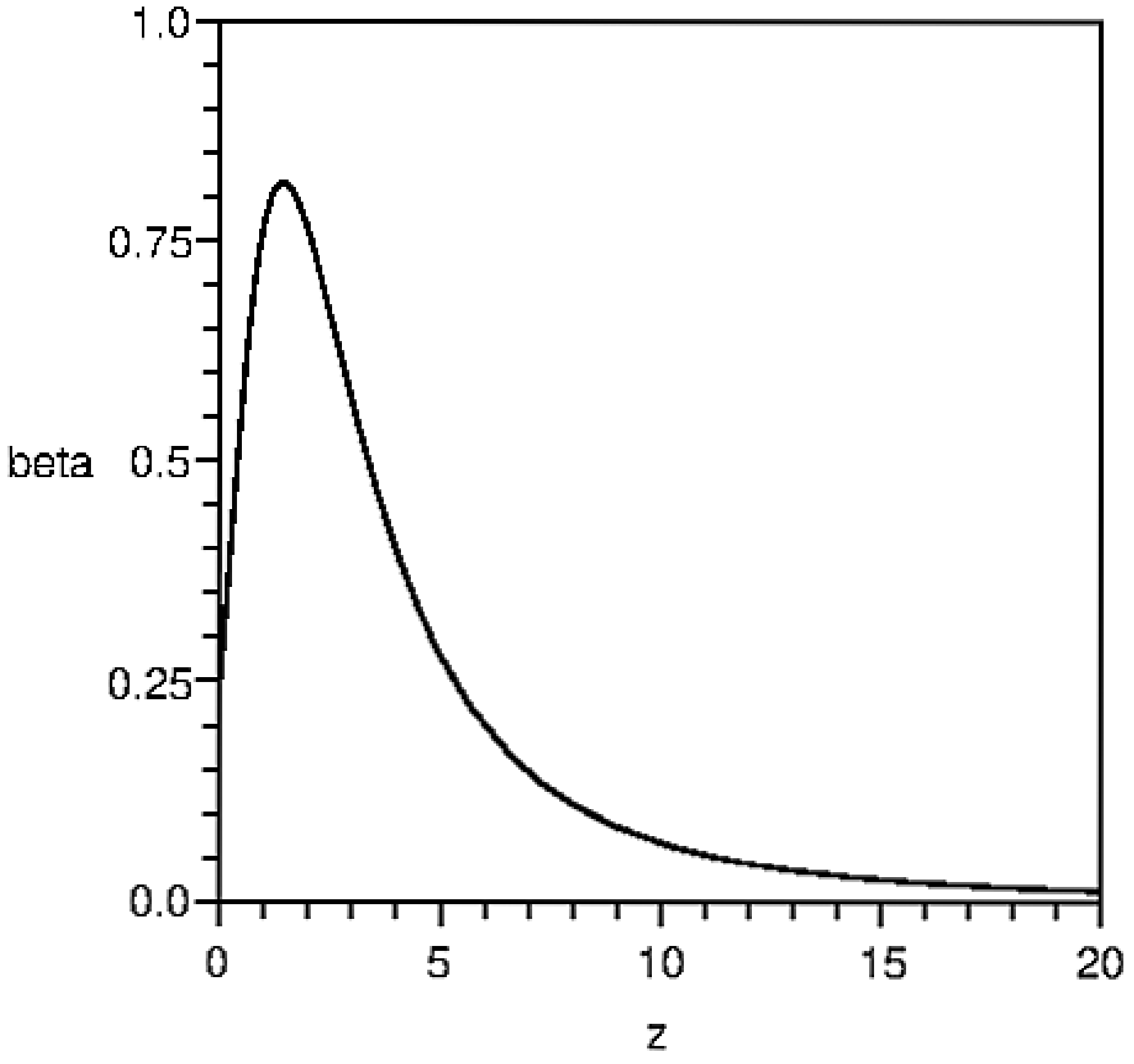

**FIG. 8.** Ratio of the energy density of cold dark matter to the energy density of the gravitational field depending on redshift.

## E. Influence of the presence of matter on the evolution of the Universe

Now we estimate the influence of matter on the evolution of the space–time manifold. In the second approximation, for a given function $\gamma(u)$, we find

$$M^{(2)}\left(u,\gamma,\frac{d\gamma}{du}\right) \cong w(u),$$

$$w(u) = 2u^{02}\left[\Omega_b \frac{f(u)}{f(u^0)} + \frac{4}{3}\Omega^* \left(\frac{f(u)}{f(u^0)}\right)^{2/3}\right] - \int_0^u \left[\Omega_b \frac{f(u)}{f(u^0)} + \frac{2}{3}\Omega^* \left(\frac{f(u)}{f(u^0)}\right)^{2/3}\right] \frac{4u^{02} du}{4u^2 - 2u + \sigma}. \quad (4.23)$$

Substituting (4.23) into (4.1), we derive the equation describing how matter in turn affects the change of the metric. The solution of this equation can be written as a quadrature:

$$\sqrt{\frac{\gamma(u)}{\gamma_{\min}}} = \psi(u) = \exp\left(\int_0^u \frac{4u du}{4u^2 - 2u + \sigma + w(u)}\right), \quad (4.24)$$

$$t - t_{st} = T\sqrt{\gamma_{\min}} \int_0^u \frac{4\psi(u) du}{4u^2 - 2u + \sigma + w(u)}. \quad (4.25)$$

The constant $\sigma$ in these relations, in the same manner as in the previous section, has to be defined together with the value of $u^0$ from a condition of the equality of the evaluated age of the Universe and the Hubble parameter to their values observed now.

$$t^0 - t_{st} = T\sqrt{\gamma_{\min}} \int_0^{u^0} \frac{4\psi(u)}{4u^2 - 2u + \sigma + w(u)} du\,,\; H^0 = \frac{1}{3T\sqrt{\gamma_{\min}}} \frac{u^0}{\psi(u^0)}. \quad (4.26)$$

The solution of this system of equations for $t^0 - t_{st} = 4.355 \times 10^{17}$ s, $H^0 = 2.181 \times 10^{-18}$ s$^{-1}$, maximum energy density $\rho_{gr\max} = 2 \times 10^{49}$ J·m$^{-3}$, and relative density of matter components: $\Omega_b = 0.0499$; $\Omega_\gamma = 5.46 \times 10^{-5}$; $\Omega_\nu^* = 1.0608 \times 10^{-4}$ has the form

$$T\sqrt{\gamma_{\min}} = 8.6912868 \times 10^{-14} \text{ s}, \; u^0 = 5.64160\ldots, \; \sigma = 0.250516884.$$

The results of calculations carried out both with and without taking into account the presence of matter in the Universe are presented in graphical form in Figs. 9–11. In view of the data provided in the previous section, it is possible to conclude that the influence of prehistory on the further course of the given dependences is insignificant in the range of redshifts less than 2.3. Furthermore, the birth of matter does not lead to a noticeable time shift in the change of the deceleration–acceleration eras.

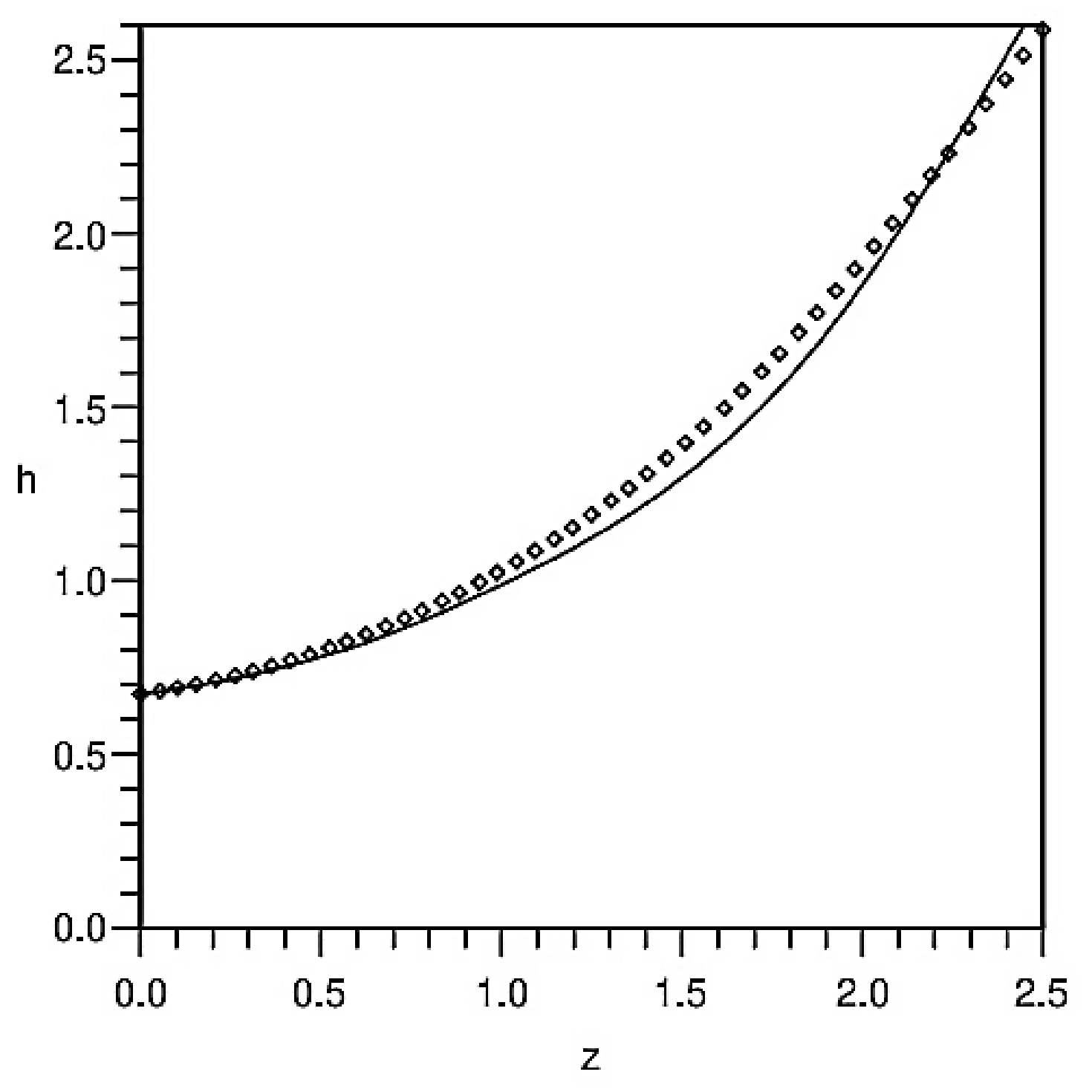

**FIG. 9.** Results of calculation of the dependence of the Hubble parameter (H=100h km·sec$^{-1}$·Mpc$^{-1}$) on red shift taking into account (points) and without taking into account (full line) the presence of matter.

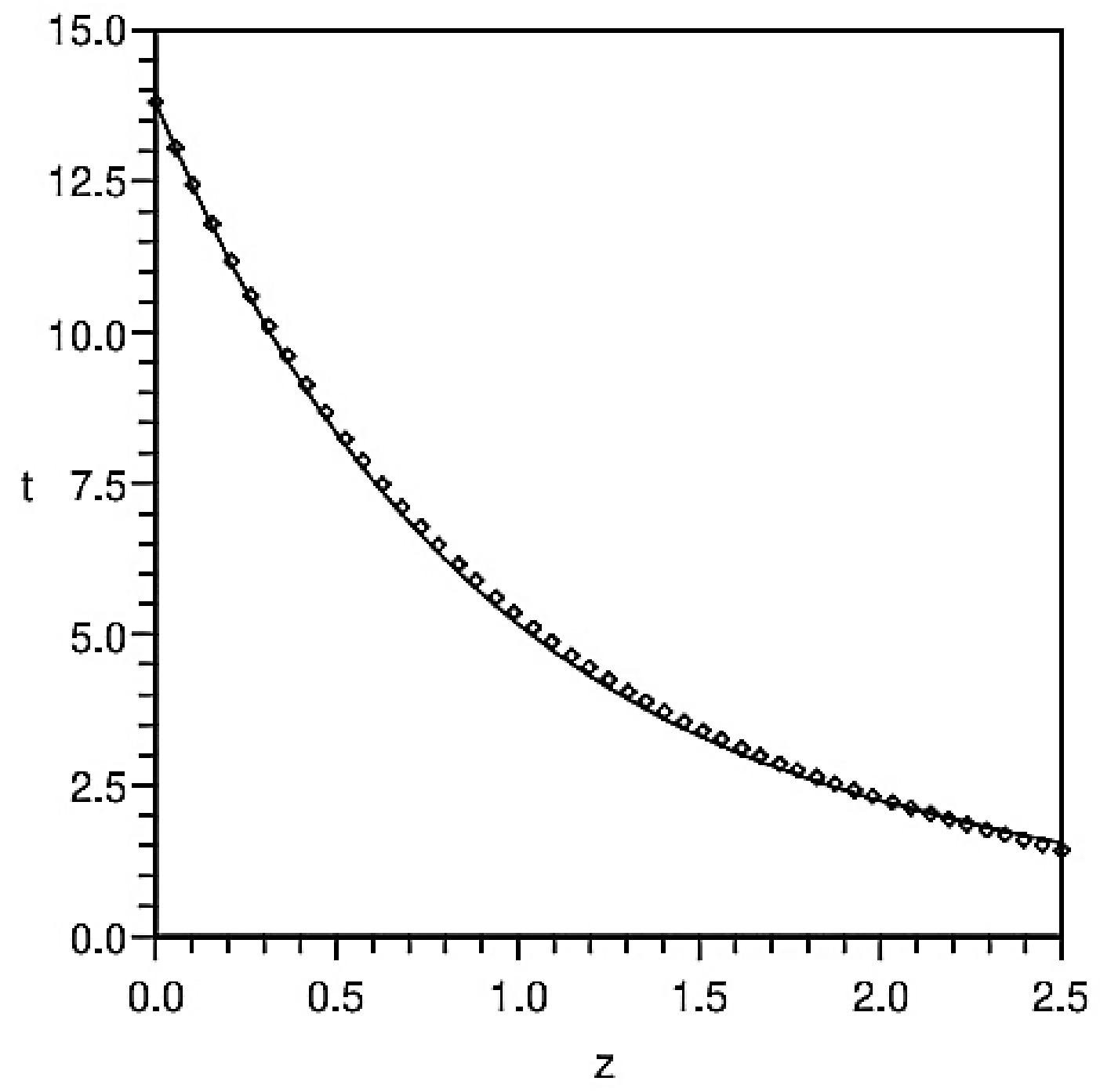

**FIG. 10.** Results of calculation of the object age (in billions of years) depending on its observed redshift taking into account (points) and without taking into account (full line) the presence of matter.

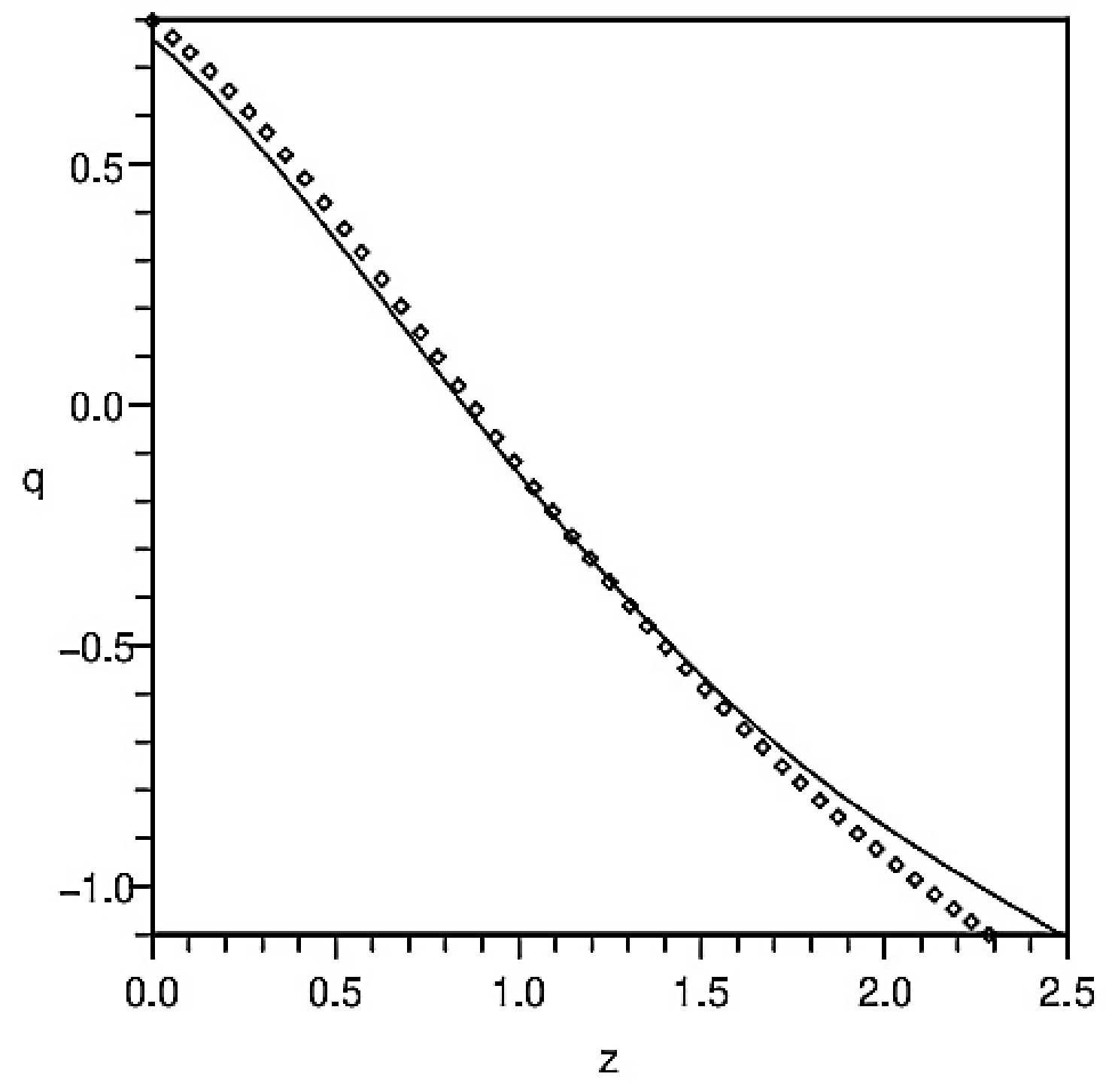

**FIG. 11.** Results of calculation of the dependence of cosmic acceleration on red shift taking into account (points) and without taking into account (full line) the presence of matter.

## V. CANONICAL QUANTIZATION OF THE THEORY OF GRAVITY WITH A CONSTRAINT

The formulation of the main provisions of a quantum theory is essentially impossible without invoking the classical theory.[23] To construct a quantum theory, it is first necessary to determine the dynamic degrees of freedom. As shown in Sec. III, in the theory of gravity with a constraint, there exists a unique homogeneous space–time with an isotropic metric of the form

$$ds^2 = g_{00}(x^0)(dx^0)^2 - \gamma^{1/3}(x^0)dx^m dx^n \delta_{mn}.$$

For this metric, the expressions for the Christoffel symbols and the nonzero components of the Ricci tensor have the form

$$\Gamma^0_{00} = \frac{1}{2g_{00}}\frac{dg_{00}}{dx^0},\ \Gamma^0_{0l} = 0,\ \Gamma^0_{nl} = \frac{1}{2g_{00}}\frac{d\gamma^{1/3}}{dx^0}\delta_{nl},\ \Gamma^m_{00} = 0,\ \Gamma^m_{0l} = \frac{1}{6\gamma}\frac{d\gamma}{dx^0}\delta^m_l,\ \Gamma^m_{nl} = 0, \quad (5.1)$$

$$R^0_0 = -\frac{1}{2\sqrt{g_{00}}}\frac{d}{dx^0}\left(\frac{1}{\gamma\sqrt{g_{00}}}\frac{d\gamma}{dx^0}\right) - \frac{1}{12g_{00}}\left(\frac{1}{\gamma}\frac{d\gamma}{dx^0}\right)^2, \quad (5.2)$$

$$R^p_k = -\frac{1}{6\sqrt{\gamma g_{00}}}\frac{d}{dx^0}\left(\frac{1}{\sqrt{\gamma g_{00}}}\frac{d\gamma}{dx^0}\right)\delta^p_k. \quad (5.3)$$

In the general case, the action of the gravitational field in the theory of gravity with a constraint has the form (2.1)

$$S_{gr} = -\frac{c^3}{16\pi G}\int (R+Q)\sqrt{-g}d^4x,\ Q = \frac{1}{\sqrt{-g}}\frac{\partial\sqrt{-g}}{\partial x^\mu}g^{\mu\nu}\frac{\partial\Phi}{\partial x^\nu}.$$

Substituting expressions for the scalar curvature and omitting the total derivatives that do not contribute to the equation of motion, we find the expression for the action of the homogeneous and isotropic space of volume $V$

$$S_{gr} = \int Ldt = \frac{c^2}{16\pi G}\int \left[\frac{1}{6}\left(\frac{1}{\gamma}\frac{d\gamma}{dt}\right)^2 + \frac{1}{\sqrt{\gamma g_{00}}}\frac{d\sqrt{\gamma g_{00}}}{dt}\frac{d\Phi}{dt}\right]\sqrt{\gamma}dt \times V, \quad (5.4)$$

where $L$ is the Lagrangian of the gravitational field, and the notation $cdt = \sqrt{g_{00}}dx^0$ is introduced.

It should be noted that in expression (5.4) we changed the sign of the action from minus to plus in comparison with (2.1). If there is a minus sign before integral (5.4), standard reasoning leads to the conclusion that this action cannot have a minimum. The correct sign is a plus.

The volume of homogeneous isotropic space can be represented in Planck units as

$$V = l^3_{pl}\lambda = \left(\frac{\hbar G}{c^3}\right)^{3/2}\lambda,\ \ \lambda = \text{const}.$$

Then, the initial action for quantization will take the form

$$S_{qgr} = \int Ldt = A\int \left[\frac{1}{6}\left(\frac{1}{\gamma}\frac{d\gamma}{dt}\right)^2 + \frac{1}{\sqrt{\gamma g_{00}}}\frac{d\sqrt{\gamma g_{00}}}{dt}\frac{d\Phi}{dt}\right]\sqrt{\gamma}\lambda dt,\ A = \frac{\hbar t_{pl}}{16\pi},\ t_{pl} = \left(\frac{\hbar G}{c^5}\right)^{1/2}. \quad (5.5)$$

In a quantum theory, it is necessary to introduce a specific value $\lambda$ for the transition to the action. This is in contrast to the classical theory, in which the equations of motion do not depend on the magnitude of the action. The solution of the gravitational equations in the theory of gravity with a constraint involves the parameter $(\gamma_{\min})^{1/2}$, which is the minimum value of the volume factor (3.21). If we take

$$\lambda = (\sqrt{\gamma_{\min}})^{-1}, \quad (5.6)$$

then action (5.5) takes the form

$$S_{qgr} = \int Ldt = A\int \left[\frac{1}{6}\left(\frac{1}{\gamma}\frac{d\gamma}{dt}\right)^2 + \frac{1}{\sqrt{\gamma g_{00}}}\frac{d\sqrt{\gamma g_{00}}}{dt}\frac{d\Phi}{dt}\right]\sqrt{\frac{\gamma}{\gamma_{\min}}}dt,\ A = \frac{\hbar t_{pl}}{16\pi}. \quad (5.7)$$

In this case, the action turns out to be scale invariant with respect to change in the magnitude of the volume factor.

Let us carry out a canonical quantization of the gravitational field based on action (5.7). We introduce the scale-invariant generalized time-dependent coordinates

$$q^1 = \sqrt{\gamma/\gamma_{\min}},\ q^2 = \ln\sqrt{g_{00}\,\gamma/\gamma_{\min}},\ q^3 = \Phi,\ 1 \le q^1 < \infty,\ -\infty < q^2,\ q^3 < \infty, \quad (5.8)$$

their velocities $v^i = \dot{q}^i$ (time derivatives), and its conjugate momenta

$$p_i = \frac{\partial L}{\partial v^i}: p_1 = A\frac{4v^1}{3q^1},\ p_2 = Aq^1v^3,\ p_3 = Aq^1v^2. \tag{5.9}$$

We find from here the velocities as functions of coordinates and momenta

$$v^1 = \frac{3q^1p_1}{4A},\ v^2 = \frac{p_3}{Aq^1},\ v^3 = \frac{p_2}{Aq^1}. \tag{5.10}$$

Let us find the energy of the gravitational field in the Lagrangian's formalism

$$E = \frac{\partial L}{\partial v^i}v^i - L = A\left(\frac{2}{3q^1}(v^1)^2 + q^1v^2v^3\right). \tag{5.11}$$

Eliminating the velocities from this relation, using (5.10) we find the Hamiltonian of the gravitational field

$$H_{gr} = \frac{1}{A}\left(\frac{3}{8}q^1(p_1)^2 + \frac{1}{q^1}p_2p_3\right). \tag{5.12}$$

We find the wave equation of the gravitational field passing in accordance with the rules of canonical quantization from coordinates and momenta to their operators,[23] and replace the product of non-commutative operators with the symmetrized product

$$i\hbar\frac{\partial\Psi}{\partial t} = \widehat{H}_{gr}\Psi,\quad \widehat{H}_{gr} = \frac{1}{A}\left(\frac{3}{16}\hat{q}^1(\hat{p}_1)^2 + \frac{3}{16}(\hat{p}_1)^2\hat{q}^1 + \frac{1}{\hat{q}^1}\hat{p}_2\hat{p}_3\right),$$

$$i\hbar\frac{\partial\Psi}{\partial t} = -\frac{3\pi\hbar}{t_{pl}}\left(q^1\left(\frac{\partial}{\partial q^1}\right)^2 + \left(\frac{\partial}{\partial q^1}\right)^2 q^1 + \frac{16}{3q^1}\frac{\partial}{\partial q^2}\frac{\partial}{\partial q^3}\right)\Psi. \tag{5.13}$$

The Hamiltonian depends on only one coordinate $q^1$. Therefore, the wave function can be represented as a superposition of products of the wave eigenfunctions of the energy $E$ and momenta $p_2$, $p_3$.

$$\Psi(t,q^i) = \int a(E)b(p_2)c(p_3)\exp\frac{i}{\hbar}(-Et + p_2q^2 + p_3q^3)\phi_E dEdp_2dp_3, \tag{5.14}$$

where

$$E\phi_E = -\frac{6\pi\hbar}{t_{pl}}\left(q^1\left(\frac{\partial}{\partial q^1}\right)^2 + \frac{\partial}{\partial q^1} - \frac{8}{3q^1}\frac{p_2p_3}{\hbar^2}\right)\phi_E. \tag{5.15}$$

Let us denote by a prime symbol the derivative with respect to the variable $x = q^1$. Then, Eq. (5.13) takes the form

$$x\phi_E'' + \phi_E' - \frac{8}{3x}\frac{p_2p_3}{\hbar^2}\phi_E + \frac{\mathrm{E}t_{pl}}{6\pi\hbar}\phi_E = 0. \tag{5.16}$$

This equation belongs to the type of equations solvable in terms of Bessel functions (Ref. 24, p. 245). If

$$w'' + \frac{1-2\alpha}{x}w' + \left[(\beta\gamma x^{\gamma-1})^2 + \frac{\alpha^2-\nu^2\gamma^2}{x^2}\right]w = 0,\quad \alpha,\beta,\gamma - \text{const},$$

then $w = x^\alpha Z_\nu(\beta x^\gamma)$, where $Z_\nu(\beta x^\gamma)$ is a Bessel function of the 1st, 2nd, or 3rd kind. Comparing the last two equations, we find

$$\alpha = 0,\quad \gamma = \frac{1}{2},\quad \beta = \pm\left(\frac{2Et_{pl}}{3\pi\hbar}\right)^{1./2},\quad \nu = \pm 4\sqrt{\frac{2p_2p_3}{3\hbar^2}},\quad \phi_E = Z_\nu(\beta\sqrt{x}),\ x = q^1. \tag{5.17}$$

The presence of the edge in the theory of gravity with a constraint means that the evolution of the metric has a beginning. From the point of view of quantum theory, this should be interpreted as the wave function equaling zero before the initial moment of time. It is necessary to take $\phi_E(q^1) = 0$ for $q^1 \le 1$ due to the requirement for the single-valuedness and continuity of the wave function across the entire space.[23]

The solutions of Eq. (5.16) will decrease for large values of the argument only if the order of Bessel functions is real. This leads to the condition $p_2 \times p_3 \ge 0$. This requires that both momenta $p_2$ and $p_3$ have either non-positive or non-negative values. In addition, solutions will be valid for $\beta \ge 0$. Thus, for $E \ge 0$, the solution to Eq. (5.16) satisfying all these conditions is the functions

$$\phi_E(q^1,E) = J_\nu\left(\sqrt{\frac{2Et_{pl}}{3\pi\hbar}}\right)N_\nu\left(\sqrt{q^1\frac{2Et_{pl}}{3\pi\hbar}}\right) - N_\nu\left(\sqrt{\frac{2Et_{pl}}{3\pi\hbar}}\right)J_\nu\left(\sqrt{q^1\frac{2Et_{pl}}{3\pi\hbar}}\right),\ \nu = \pm 4\sqrt{\frac{2p_2p_3}{3\hbar^2}}. \tag{5.18}$$

For E <0, it is impossible to construct a solution that would vanish for $q^1 = 1$ and, at the same time, would be bounded at infinity. *Therefore, the gravitational field has only a continuous spectrum of energy, the spectrum of discrete energy levels is absent.*

Using the expression of Neumann functions via Bessel functions[24]

$$N_\nu(x) = \frac{1}{\sin\nu\pi}\,[J_\nu(x)\cos\nu\pi - J_{-\nu}(x)] \quad (\nu \neq n),$$

we write $\phi_E(q^1, E)$ (if ν is not an integer) as follows:

$$\phi_E(q^1,E) = \frac{1}{\sin\nu\pi}\left[J_{-\nu}\left(\sqrt{\frac{2Et_{pl}}{3\pi\hbar}}\right)J_\nu\left(\sqrt{q^1\frac{2Et_{pl}}{3\pi\hbar}}\right) - J_\nu\left(\sqrt{\frac{2Et_{pl}}{3\pi\hbar}}\right)J_{-\nu}\left(\sqrt{q^1\frac{2Et_{pl}}{3\pi\hbar}}\right)\right], \nu = \pm 4\sqrt{\frac{2p_2p_3}{3\hbar^2}}$$

Hence it follows that this solution is an even function of ν and the energy levels are not degenerate.

The general view of the wave function of the very early Universe has the form of (5.14) (in the presence of only the gravitational field, that is, before the formation and separation of matter), and $\phi_E(q^1, E)$ is given by the relations (5.18). To define the specific wave function, it is necessary to set it at the initial time $t = t_{st}$ (to simplify the notation, we will consider it equal to zero). Without such information, we can go a little further based on the results of the classical theory. According to (3.17) and (3.13), the "velocities" are related to each other by the relations

$$\frac{1}{\sqrt{\gamma g_{00}}}\frac{d\sqrt{\gamma g_{00}}}{dt} = \frac{1}{2T\sqrt{\gamma}},\;\; T = \text{const},\;\; \frac{dq^2}{dt} = \frac{1}{2T\sqrt{\gamma_{\min}}q^1}, \qquad (5.19)$$

$$\frac{d\Phi}{dt} = -\frac{1}{3\gamma}\frac{d\gamma}{dt} + \frac{\sigma}{3T\sqrt{\gamma}},\;\; \sigma = \text{const},\;\; \frac{dq^3}{dt} = -\frac{2}{3q^1}\frac{dq^1}{dt} + \frac{\sigma}{3q^1T\sqrt{\gamma_{\min}}}. \qquad (5.20)$$

From (5.19), taking into account (5.9), it follows that at $t = 0$ the momentum $p_3$ had a certain positive value

$$p_3^0 = \frac{\hbar t_{pl}}{32\pi T\sqrt{\gamma_{\min}}} > 0. \qquad (5.21)$$

Similarly, from (5.20), taking into account (5.9) and $v^1(0) = 0$, it follows that

$$p_2^0 = -\frac{\hbar t_{pl}}{24\pi}v^1(0) + \frac{\hbar t_{pl}\sigma}{48\pi T\sqrt{\gamma_{\min}}},\; p_2^0 = \frac{\hbar t_{pl}\sigma}{48\pi T\sqrt{\gamma_{\min}}} > 0. \qquad (5.22)$$

The momentum $p_2$ also had a certain positive value, and the product of the momenta values satisfies the condition $p_2^0 \times p_3^0 > 0$. Thus, the wave function can be represented at the initial moment of time as

$$\Psi(0,q^i) = \psi(q^1)\exp i\frac{t_{pl}(2\sigma q^2 + 3q^3)}{96\pi T\sqrt{\gamma_{\min}}},\;\; \nu^0 = 4\sqrt{\frac{2p_2^0p_3^0}{3\hbar^2}} = \frac{t_{pl}\sqrt{\sigma}}{12\pi T\sqrt{\gamma_{\min}}} << 1. \qquad (5.23)$$

The square of the function $\psi(q^1)$ gives the probability density of different values of the coordinate $q^1$ for $t = 0$. The function $\psi^2(q^1)$ must be continuous, equal to zero for $q^1 = 1$, and exponentially decrease at infinity. These conditions are satisfied by the gamma distribution of the probability density[25] (with two parameters and a shift along the coordinate).

$$\psi^2(q^1) = \frac{(q^1-1)^k}{\Gamma(k)\mu^{k+1}}\exp\left[-\frac{(q^1-1)}{\mu}\right],\quad k > 0,\quad q^1 \geq 1. \qquad (5.24)$$

Let us assume that $k = 1$ for eliminating the random distributions with extreme values of the derivative *0* and ∞ (as non-physical) at $q^1 = 1$. In this case, the average deviation of $q^1$ from unity is $2\mu$, and the variance of the distribution is $2\mu^2$.

Let us assume that the root-mean-square fluctuation of coordinate values is proportional to the ratio of the Planck time to the characteristic time of the theory of gravity with a constraint. This ratio is included in the wave function as the index $\nu = \nu^0$ (5.23) of a Bessel function, therefore

$$\sqrt{2}\mu = \nu^0,\; \nu^0 = \frac{t_{pl}\sqrt{\sigma}}{12\pi T\sqrt{\gamma_{\min}}} << 1. \qquad (5.25)$$

Thus, the initial wave function (5.23) is completely defined, and along with it, in principle, the wave function of the very early Universe is also defined. Indeed, it follows now from (5.14), (5.18), (5.23), (5.24), and (5.25) that

$$\frac{\sqrt{2(q^1-1)}}{\nu}\exp\left[-\frac{(q^1-1)}{\sqrt{2}\nu}\right] = \int_0^\infty a(E)\phi_E(q^1,E)dE\,,\ \nu=\nu^0. \qquad (5.26)$$

From a physical point of view, this integral equation specifies the decomposition of $\psi(q^1)$ with respect to eigenfunctions of the Hamiltonian operator. Let us consider the solution of this equation.
We multiply both sides of the equation by $\phi_E(q^1,E)$ and integrate with respect to $q^1$ from 1 to $\infty$.

$$\int_1^\infty \frac{\sqrt{2(q^1-1)}}{\nu}\exp\left[-\frac{(q^1-1)}{\sqrt{2}\nu}\right]\phi_E(q^1,E)dq^1 = \int_1^\infty\left(\int_0^\infty a(E')\phi_{E'}(q^1,E')dE'\right)\phi_E(q^1,E)dq^1. \quad (5.27)$$

By virtue of Weber's integral theorem (Ref. 26 Ch. XIV, Sect. 14.52, eq. 6), the multiple integral in (5.27), after appropriate re-designations, can be represented as

$$\int_1^\infty\left(\int_0^\infty a(E')\phi_{E'}(q^1,E')dE'\right)\phi_E(q^1,E)dq^1 = \frac{6\pi\hbar}{t_{pl}}\left[J_\nu^2\left(\sqrt{\frac{2t_{pl}E}{3\pi\hbar}}\right)+N_\nu^2\left(\sqrt{\frac{2t_{pl}E}{3\pi\hbar}}\right)\right]a(E). \quad (5.28)$$

From (5.27) and (5.28) it follows that

$$a(E) = \frac{t_{pl}}{6\pi\hbar}\left[J_\nu^2\left(\sqrt{\frac{2t_{pl}E}{3\pi\hbar}}\right)+N_\nu^2\left(\sqrt{\frac{2t_{pl}E}{3\pi\hbar}}\right)\right]^{-1}\int_1^\infty \frac{\sqrt{2(q^1-1)}}{\nu}\exp\left[-\frac{(q^1-1)}{\sqrt{2}\nu}\right]\phi_E(q^1,E)dq^1. \quad (5.29)$$

The wave function of the very early universe is now fully defined

$$\Psi(t,q^i) = \int_0^\infty a(E)\exp i\left[\frac{t_{pl}(2\sigma q^2+3q^3)}{96\pi T\sqrt{\gamma_{\min}}}-\frac{Et}{\hbar}\right]\phi_E(q^1,E)dE\,,\ \nu=\nu^0. \qquad (5.30)$$

## VI. CONCLUSIONS

The theory of gravity with a constraint, as the canonical theory, is based on the Hilbert action. Within the framework of the model proposed by the author, the fundamental differences from the standard cosmological model in a description of the evolution process of the Universe are as follows:

- the constraint defines an edge with a zero-world physical anisotropic time at the restriction of the group of admissible coordinate transformations;
- the gravitational field is endowed with all the properties of a material medium: energy, pressure, entropy and temperature;
- it is possible to construct a space-time manifold, in which only its boundary is singular;
- from the classical point of view, the process of the evolution of the Universe begins from a state with a minimum nonzero value of the scale factor and equal to zero energy;
- by virtue of the definition of an energy–momentum density tensor adopted in the paper, the pressure of the gravitational field at the initial moment turns out to be negative, as a result of which the growth of the scale factor begins. At the same time, the energy density of the gravitational field also grows in proportion to the growth rate squared of the volume factor. This process has an avalanche-like character ("big bang") and will continue until the energy density reaches its maximum value and begins to decrease due to the energy consumption for the adiabatic expansion of the Universe;
- Despite the presence of a singularity at the boundary, the described classical model of the evolution of the Universe allows the construction of a canonical (or using path integrals) quantum theory of the gravitational field on its basis. The wave function of the very early Universe has been constructed.
- The available experimental data on the temperature of the CMB radiation allow us to conclude that the maximum global energy density in the Universe has never exceeded $1\times10^{50}$ J·m$^{-3}$ ~ $(1.5\ \text{TeV})^4$, the maximum temperature of the matter fields has not exceed $1.230\times10^{11}$ K, and the relative energy density of neutrinos is currently less than $1.061\times10^{-4}$.

- The global energy density of the Universe is currently 94.5% composed of the energy density of the gravitational field, and all known types of matter only contribute 5.5%.
- The accuracy of the available astronomical observations is still insufficient to choose between the predictions of GR and the proposed theory of gravity. However, over the past twenty years, the physical natures of dark energy, dark matter, and inflatons have not been established, and no new particles with suitable properties have been detected at the LHC. This is an essential argument for doubting their existence.

From the point of view of the theory presented here, all observable effects associated with dark energy and dark matter are only manifestations of the material essence of the gravitational field. On the one hand, in the present era of the second acceleration, the gravitational field has a negative pressure; that is, it behaves like hypothetical dark energy. On the other hand, the energy density of the gravitational field exceeds the average energy density of matter on the large-scale structure of the Universe. This energy, which is not taken into account within the framework of GR and has properties attributed to dark matter, can contribute to an increase in the speed of the observed gravitationally bound objects. In addition, the pressure of the gravitational field was negative in the very early Universe also and, as mentioned above, already within the framework of the classical approach at zero initial energy density, this leads to the Big Bang, therefore there is no need for a hypothesis about the existence of any inflatons.